\documentclass[twocolumn]{aastex62}
\usepackage[toc]{appendix}
\usepackage{amsmath}
\usepackage{hyperref}
\usepackage{tabularx}
\usepackage{graphicx}
\usepackage{comment}
\usepackage{subfigure}

\shorttitle{High-frequency wave power in the chromosphere}
\shortauthors{Molnar et al.}
\begin{document}

\title{High-frequency wave power observed in the chromosphere with IBIS and ALMA}

\correspondingauthor{Momchil Molnar}
\email{momo@nso.edu}

\author[0000-0003-0583-0516]{Momchil E. Molnar}
\altaffiliation{DKIST Ambassador}
\affil{Department of Astrophysical and Planetary Sciences,
	    Laboratory for Atmospheric and Space Physics,
	    University of Colorado, Boulder, CO, USA}
\affil{National Solar Observatory, Boulder, CO, USA}

\author[0000-0001-8016-0001]{Kevin P. Reardon}
\affil{National Solar Observatory, Boulder, CO, USA}
\affil{Department of Astrophysical and Planetary Sciences,
	    Laboratory for Atmospheric and Space Physics,
	    University of Colorado, Boulder, CO, USA}

\author[0000-0002-3699-3134]{Steven R. Cranmer}
\affil{Department of Astrophysical and Planetary Sciences, 
       Laboratory for Atmospheric and Space Physics,
 	   University of Colorado, Boulder, CO, USA}

\author[0000-0001-7458-1176]{Adam F. Kowalski}
\affil{Department of Astrophysical and Planetary Sciences, 
       Laboratory for Atmospheric and Space Physics,
       University of Colorado, Boulder, CO, USA}
\affil{National Solar Observatory, Boulder, CO, USA}

\author{Yi Chai}
\affil{Center for Solar-Terrestrial Research, New Jersey Institute of Technology,
	    Newark, NJ 07102, USA}

\author[0000-0003-2520-8396]{Dale Gary}
\affil{Center for Solar-Terrestrial Research, New Jersey Institute of Technology,
	    Newark, NJ 07102, USA}

\begin{abstract}

We present observational constraints on
the chromospheric heating contribution from acoustic
waves with frequencies between
5 and 50 mHz.
We utilize observations from the Dunn Solar Telescope
in New Mexico complemented with
observations from the Atacama Large Millimeter Array
collected on 2017 April 23. The properties of the power spectra of the
various quantities are derived from
the spectral lines of \ion{Ca}{2} 854.2 nm, \ion{H}{1}
656.3 nm, and the millimeter continuum at 1.25 mm and 3 mm.
At the observed frequencies the diagnostics almost
all show a power law behavior, whose particulars (slope, peak and white noise
floors) are
correlated with the type of solar feature (internetwork, network, plage).
In order to disentangle the vertical versus transverse plasma motions 
we examine two different fields of view;
one near disk center and the other close to the limb.
To infer the acoustic flux in the middle chromosphere, we compare
our observations with synthetic observables from the
time-dependent radiative hydrodynamic RADYN code.
Our findings show that
acoustic waves carry up to about 1 kW~m$^{-2}$ of
energy flux in the middle chromosphere, which is not enough to maintain the
quiet chromosphere, contrary to previous publications.
\end{abstract}

\vspace{.2cm}

\section{Introduction}  \label{sec:Introduction}

Balancing the radiative losses of the non-magnetic chromosphere requires an energy input of
about 4 kW~m$^{-2}$
\citep{1977ARA&A..15..363W}.
The two most widely accepted theoretical
frameworks for chromospheric heating are the same as for the corona:
wave dissipation or ubiquitous small scale magnetic
reconnection~\citep{2019ARA&A..57..189C}.
There is evidence for both mechanisms being at work
in the chromosphere, but
definitive observations constraining their relative importance
are still elusive. In this paper we concentrate on
quantifying the contribution from compressive waves in the chromosphere which produce
measurable Doppler shifts in chromospheric
diagnostics.

The first to suggest that the chromosphere
can be kept in its thermal state by dissipation of acoustic waves
were \cite{1946NW.....33..118B} and \cite{Schwarzschild1948}.
They suggested that the
convective overshoot at the boundary of the the upper convective zone and the photosphere
drives acoustic waves with a fairly broad range of periods.
Those waves with frequencies above the acoustic cutoff frequency
(roughly 5 mHz or 200 seconds) can propagate upward into the
chromosphere \citep{1974soch.book.....B}. As the acoustic waves
move into higher layers in the solar atmosphere,
they find a strongly decreasing plasma density while the temperature, and hence sound speed, remains (almost) constant, which results in
amplitude growth and wave steepening.
Waves that steepen into shocks can
dissipate their energy and supply the heat needed to maintain
the chromospheric plasma in its basal state \citep{1992ApJ...397L..59C}.
Intermittent shocks would ionize the hydrogen and helium in the chromosphere
out of equilibrium
and maintain the chromospheric ionization state away from statistical
equilibrium due to the long recombination timescales
\citep{2002ApJ...572..626C}.

The magnetic field in the chromosphere
allows for the existence of magnetosonic wave modes and the similarity between
the sound and Alfv\`en speeds allows the easy conversion
between them \citep{2008SoPh..251..251C}.
The multitude of magneto-acoustic wave modes and
their numerous damping mechanisms
expands the possible wave propagation
scenarios, but does not alter the basic premise of the
theory of how energy is being transported from the
convection zone to the chromosphere.
For a recent review on the subject of magnetosonic waves, the reader
should see \cite{2015SSRv..190..103J}.

The observational evidence constraining the
energy contribution of wave heating in the chromosphere has been
inconclusive. \cite{2002A&A...395L..51W}, using 
Fabry-Perot imaging spectroscopy of the \ion{Fe}{1} 543.4
nm line, inferred around 0.9 kW~m$^{-2}$
acoustic flux from waves with periods
between 50 s and 100 s (10-20 mHz) at height of 600 km above the photosphere.
They estimated that the large extent in height
of the velocity response functions reduced their observed wave 
amplitudes by a factor of three, and arrived at an actual acoustic flux of 3 kW~m$^{-2}$.
On the other hand, \cite{2005Natur.435..919F} used TRACE 
intensity-only observations of the 160 nm UV continuum (sampling
the upper photosphere at height of 450 km),
coupled with
self-consistent simulations of the solar atmospheric oscillations,
to derive an acoustic wave flux in the frequency interval
5-28 mHz of 0.4 kW~m$^{-2}$. This would not be sufficient to sustain the non-magnetic
chromosphere, but other authors
\citep[e.g.,][]{2007ApJ...657L..57C, 2007ASPC..368...93W}
have argued that the limited angular resolution of the TRACE observations
and other model assumptions lead to significant underestimation of the wave flux in that analysis.

Measurements of velocities in two photospheric spectral lines with
higher temporal and spatial resolution by
\cite{2009A&A...508..941B, 2010ApJ...723L.134B}
showed the presence, after taking into account the width of 
the velocity response functions, of significant acoustic flux
(up to 3.8 kW~m$^{-2}$) in the middle photosphere between 5 and 15 mHz.
More recent work by \cite{2020ApJ...890...22A,  2020arXiv200802688A}
utilizing  observations of the chromospheric \ion{Ca}{2} 854.2 nm
and \ion{H}{1} Balmer-$\alpha$ and Balmer-$\beta$ lines with a scanning spectrograph
reached similar conclusions of about 5 kW~m$^{-2}$ flux in the chromosphere.
However, the latter authors did not account for the width of the velocity 
response function \citep{1980A&A...84...96M}, which makes their flux estimation a lower
bound for the actual wave energy flux.

Another still poorly understood aspect of the wave heating theory
is the contribution from high-frequency waves, above 30 mHz.
The sparsity of observations in this regime has been due to the difficulty
of obtaining high-temporal-cadence spectral information
of the chromosphere at the high spatial resolution
required to resolve the small-scale chromospheric structures.
In one of the few studies in this regime, \cite{2000A&A...360..742H}
has shown intermittent wavelet power
up to 50 mHz in upper-chromospheric and transition-region lines
taken with SUMER on the SOHO spacecraft.

Some of the aforementioned studies 
take into account the attenuation of the observed wave amplitudes 
 when the contribution to the observed spectral signal becomes similar to or greater than the wavelength of the propagating waves~\citep{1980A&A...84...96M}. 
Consideration of this effect is essential for inferring the flux, as it has been estimated to be a factor of between 2 to 10 for frequencies 
between 5 and 50 mHz \citep{2002A&A...395L..51W, 2009A&A...508..941B}. 
The latter authors used a static
semi-empirical 1D model to calculate the wave 
response in photospheric lines using a perturbative approach with sinusoidal 
waves. 
This approach is not similarly 
applicable in the
chromosphere, where the waves are generally not sinusoidal and could be strongly
affected by radiative losses.
A more realistic approach is undertaken by
\citet{2005Natur.435..919F} and 
\citet{2007ASPC..368...93W} who use 
time-dependent hydrodynamic simulations 
to infer the actual wave attenuation including the
effect of radiative wave damping. We show in 
Section~\ref{sec:RADYN_results} that this approach
is better as
it naturally explains the high frequency signal in
our observations. Furthermore, we argue that 
modeling based on semi-empirical 1D atmospheres
might be overestimating the contributions from the high-frequency 
waves. 
\cite{2008ApJ...683L.207R} showed that Doppler diagnostics in the chromosphere have a power-law behavior from the acoustic cutoff frequency out to 20 mHz. It is not understood whether this trend is due to the true distribution of acoustic oscillations at these frequencies, or the result of a frequency-dependent attenuation of the chromospheric signal.

This paper presents observations of high frequency wave Doppler velocity signal
in the solar chromosphere. We obtained a data set of cotemporal
observations with the Interferometric Bidimensional Imaging Spectrograph (IBIS)
instrument at the
Dunn Solar Telescope (DST) and with the Atacama Large Millimeter
Array (ALMA) in a sparsely explored
temporal regime up to 50 mHz. In Section~\ref{sec:High_frequency_power}
we present evidence for
the presence of high-frequency power in the
velocity measurements of the H$\alpha$ and \ion{Ca}{2} 
854.2 nm lines as well in the
ALMA brightness temperatures. To infer the
wave energy fluxes from our observations
we model the propagation of acoustic waves throughout the solar
atmosphere with the RADYN code
in Section~\ref{sec:Transmission_function}. The results from the
modeling are presented
in Section~\ref{sec:RADYN_results} and the discussion of our results and the
conclusions are summarized
in Section~\ref{sec:Conclusions}.

\section{Observations and Data Processing}
\label{sec:Observations}

\begin{figure*}
	\includegraphics[width=\textwidth]{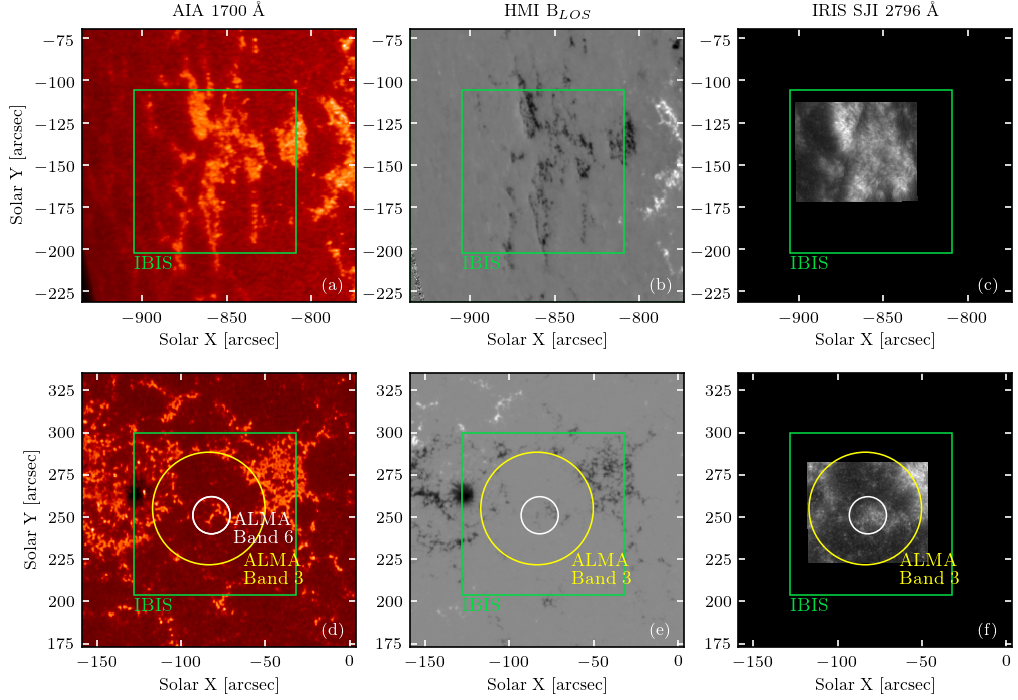}
	\caption{Context images of the observed fields of view. The top row
		(panels \emph{a, b} and \emph{c}) are FOV 1 (as seen at
		14:25 UT) and the bottom row (panels \emph{d, e} and \emph{f})
		 are FOV 2 (as seen at 15:06 UT).
		\textbf{\textit{Left panels}}: AIA 1700 {\AA} image showing photospheric emission; 
		\textbf{\textit{Central panels}}: HMI
		Line-of-sight (LOS) magnetogram, where black and white denote high magnetic flux, gray areas
		depict close to zero magnetic flux; 
		\textbf{\textit{Right panels}}: IRIS slitjaw image at 2796 {\AA}.
		The field of view of IBIS is shown as the green rectangle; the FOV of ALMA
		Band 3 is drawn as the yellow circle
		and the FOV of ALMA Band 6 as the white circle.}
	\label{fig:context_1}
\end{figure*}

We obtained coordinated solar observations
with ALMA~\citep{2009IEEEP..97.1463W} and the DST \citep{1964ApOpt...3.1353D,1991AdSpR..11..139D}
on 2017 April 23.
These observations were part of
ALMA Project ID 2016.1.01129S/Cycle 4. See \cite{2019ApJ...881...99M}
for some of the initial results from this observational campaign.

Two separate fields of view (FOV) were observed on 2017 April 23
and they are referred chronologically 
as FOV 1, near the limb, and FOV 2, close to disk center.
The targets in both cases were regions of magnetic 
network or plage, but they were observed at different inclination 
angles to the solar surface, which is essential for the 
discussion in Section~\ref{subsec:FOV1_power}.
Context images of FOV 1 and 2 are provided from
SDO/AIA~\citep{2012SoPh..275...17L},
SDO/HMI~\citep{2012SoPh..275..229S},
and IRIS~\citep{2014SoPh..289.2733D}
in Figure~\ref{fig:context_1}. Further analysis
of the IRIS dataset is left for a forthcoming publication.

FOV 1 was observed with the DST at 13:50--15:14 UT
at solar coordinates E 66.2$^{\circ}$, S 09.8$^{\circ}$, at an inclination
 of $\mu=0.41$ (the cosine of the angle between the line 
 of sight and the solar surface normal). FOV 1 was centered on
the trailing edge of NOAA Active Region (AR) 12653.
Based on the context imaging (top panel of Figure~\ref{fig:context_1})
FOV 1 contains some active region plage
as well as internetwork regions with little magnetic field.
The inclined viewing angle results in more confusion among features in the field of view because of projection effects of predominantly vertical features and longer integration along the line of sight. 


FOV 2 was observed with the DST at
15:15--18:19 UT at solar
coordinates E 4.9$^{\circ}$, N 10.9$^{\circ}$
at an inclination of $\mu=0.98$.
FOV 2 was centered on the leading edge of the active
region NOAA AR 12651. 
There is a magnetic concentration in the center of the field, surrounded by a
largely field-free internetwork area, especially in the southern portion of the field.
There is a region of plage in the northwestern corner of the field and
the leading edge of the penumbra/superpenumbra on the northeastern corner.

\subsection{DST observations}
\label{subsec:DST_observations}

The DST took observations on 2017 April 23
between 14:00 UT and 18:40 UT, under conditions of excellent to good seeing.
The instrument setup
included
IBIS~\citep{Cavallini2006,2008A&A...481..897R}, the Facility InfraRed
Spectrograph \citep[FIRS,][]{FIRS}
and the Rapid Oscillations in the Solar Atmosphere 
instrument \citep[ROSA, ][]{2010MmSAI..81..763J},
which provided thorough coverage of key spectral lines in the optical and
the near-IR parts of the spectrum. All of the
instruments were fed by the high-order adaptive optics
system~\citep{2004SPIE.5490...34R}. None of the instruments were
run in polarimetric mode, as high temporal cadence was the
priority for this study.

\begin{table}[]
  \begin{center}
\begin{tabular}{c c c c}

\underline{Time [UT]} & 
\underline{Spectral Interval} &
\underline{Cadence [s]} / \ &
\underline{Target} \\

 & & \underline{Number of scans} & \\ 
14:13--14:18 & Ca II 854.2 nm & 3.13 / 77 & FOV 1  \\
15:13--15:18 & Ca II 854.2 nm  & 3.11 / 100 & FOV 2 \\
15:18--15:28 & H$\alpha$ 656.3 nm & 3.68 / 150 & FOV 2 \\
15:39--15:46 & Ca II 854.2 nm & 3.28 / 120 & FOV 2 \\
15:53--16:01 & ALMA 1.25 mm & 2.0 / 238 & FOV 2 \\
16:03--16:11 & ALMA 1.25 mm & 2.0 / 238 & FOV 2 \\
17:04--17:11 & Ca II 854.2 nm & 3.27 / 120 & FOV 2 \\
17:11--17:19 & H$\alpha$ 656.3 nm & 3.67 / 120 & FOV 2\\
17:19--17:29 & ALMA 3 mm & 2.0 / 300 & FOV 2 \\
17:31--17:41 & ALMA 3 mm & 2.0 / 300 & FOV 2\\
\end{tabular}
\caption{Observations used in this work.}
\label{table:Observations}
  \end{center}
  \end{table}

\begin{figure}
	\includegraphics[width=.45\textwidth]{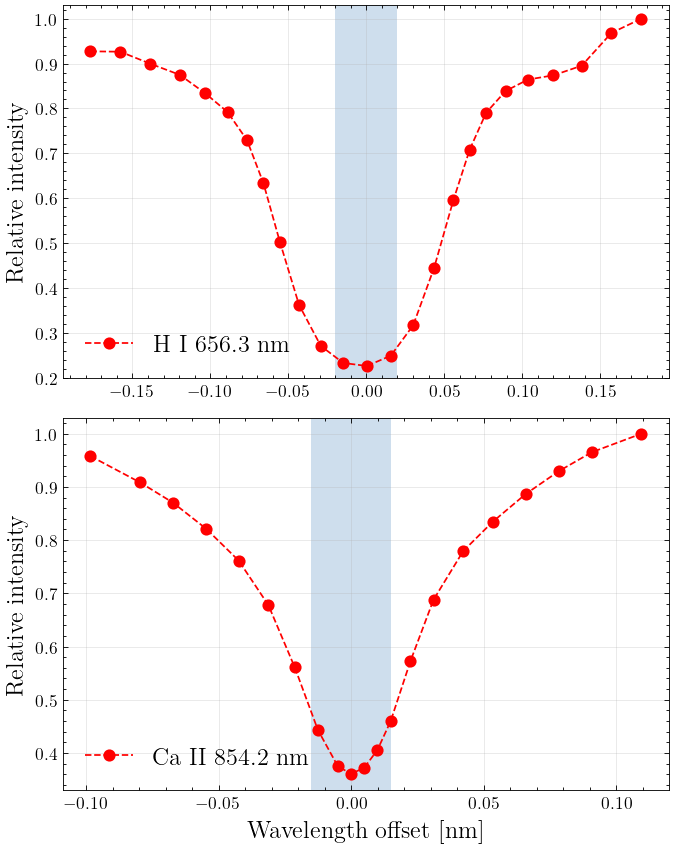}
	\caption{Average spectral profiles from the fast scans
		          with IBIS in the H$\alpha$ (656.3 nm) and \ion{Ca}{2} 854.2 nm
		          lines. The line cores were
	              sampled more densely to allow for more accurate determination of
	              the line properties derived from the line cores (velocity,
	              intensity and width). 
	              The blue shaded regions are used for the line core fitting with a parabola after a resampling on a finer (0.005 nm) grid.}
    \label{fig:Fast_scan_spec_profiles}
\end{figure}

\begin{figure*}
\includegraphics[width=\textwidth]{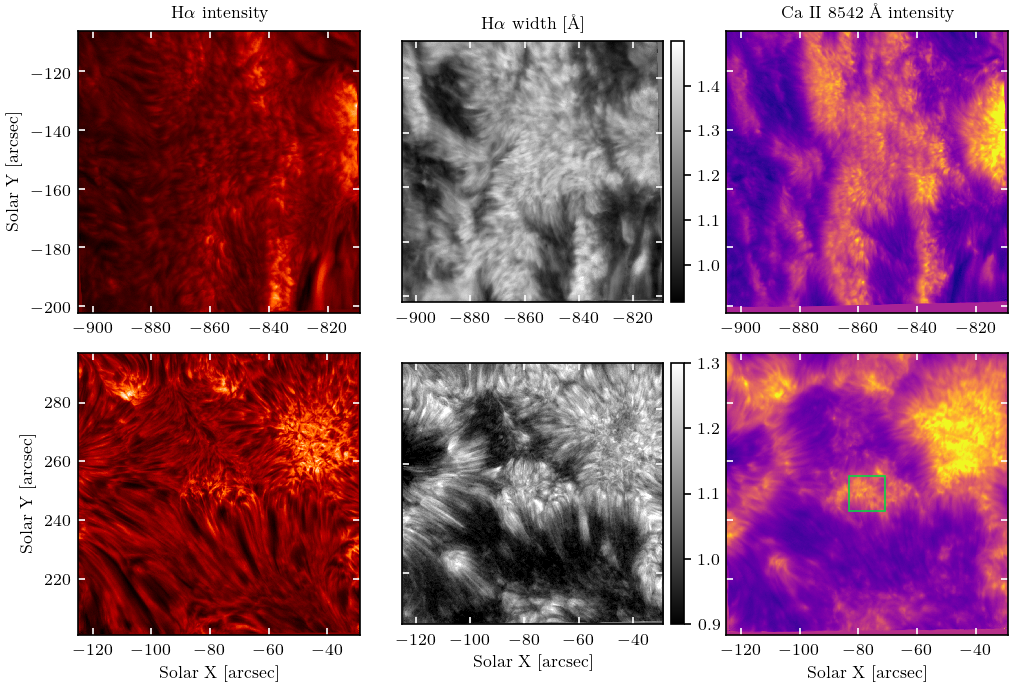}
  \caption{The fields of view observed with IBIS in the following chromospheric diagnostics:
  	\textbf{\textit{Left panels}}: H$\alpha$ 656.3 nm line core intensity;
  	\textbf{\textit{Middle panels}}: H$\alpha$ line
  	core width (defined in Section~\ref{subsec:DST_observations});
  \textbf{\textit{Right panels}}: 
  \ion{Ca}{2} 854.2 nm line core intensity.
  The top row corresponds to FOV 1 observed around 14:18 UT
  and the bottom row corresponds to FOV 2 observed at 17:20 UT.
  FOV 1 was observed close to the limb
  at $\mu$=0.41 which shortens the projection of the solar 
  features in a direction away from disc center. FOV 1
  covers mostly a plage region, while FOV 2 consists of 
  network and internetwork with some plage
  in the top right. The green rectangle in the last 
  panel shows the area from which the power spectra
  were used in Figure~\ref{fig:Power_spectra_overview}.}
  \label{fig:IBIS_FOV}
\end{figure*}

IBIS observed the spectral lines of \ion{H}{1} Balmer-$\alpha$ 656.3 nm
(H$\alpha$), \ion{Ca}{2} 854.2~nm
and \ion{Na}{1} D$_1$ 589.6~nm with an average
plate scale of 0.096\arcsec~pixel$^{-1}$.
Each scan of a single 
spectral line scan took between 3 and 4 seconds with
an overhead of about 1.5 seconds for changing the prefilters.
We utilized two different scanning strategies during the observations. 
At the beginning of each ALMA observing block at a given pointing 
and frequency band, we ran ``fast'' scans of H$\alpha$ and 
\ion{Ca}{2} 854.2 nm. Each scan consisted of 
25 and 21 wavelength points, respectively and the average profiles from 
those scans are shown in Figure~\ref{fig:Fast_scan_spec_profiles}.
Scanning a single line avoids the overhead for changing IBIS prefilters, resulting in a cadence of about 3.5 seconds for a single spectral scan
(the precise cadences are listed in Table~\ref{table:Observations}).
This observing strategy was adopted to
closely match the ALMA two-second 
sampling rate and allow
us to study the high frequency wave regime
(with Nyquist sampling of 130 mHz).
These 
scans were intended to capture the 
fast dynamics of the chromosphere,
at high temporal frequencies rarely explored 
with wide-field, bi-dimensional spectroscopy.
Table~\ref{table:Observations} contains a summary of the 
single line (``fast'') scans with IBIS and the 
ALMA time series used in this study.

Following these fast scans, we ran longer durations of
a ``standard'' 
repeating cycle of all the three 
spectral lines.
The standard scans consisted of 90
total spectral points among the three lines and had duty cycle of about 15 seconds.
This type of scan
thoroughly covers the solar atmospheric layers from the photosphere through
the middle chromosphere, making it well suited
for studying the propagation of wave energy
through the lower solar atmosphere.

To remove the seeing distortions in our narrowband
images we relied
on destretching the cotemporal 
broadband images (from the white-light
IBIS channel)
 to HMI white-light images using
the sub-aperture cross-correlation method introduced by
\cite{1988ApJ...333..427N}. We applied the destretch maps
to the
cotemporaneous narrowband images and the
resulting stability of the destretched
images was on the order of one IBIS pixel ($\sim 0.1$\arcsec).
The wavelength samplings for the fast scans
of H$\alpha$ and the \ion{Ca}{2} 854.2 nm lines are presented
in Figure~\ref{fig:Fast_scan_spec_profiles}. The lines were sampled
non-equidistantly to ensure better coverage of the line core,
which is used for the velocity, intensity and width measurements. The 
maps of the observed line core intensity and width in H$\alpha$ and 
\ion{Ca}{2} 854.2 nm for both FOVs observed with IBIS
are presented in Figure~\ref{fig:IBIS_FOV}.

Following~\cite{2009A&A...503..577C}, we measured the width of the chromospheric line cores as the separation of the two wing positions at which the intensity reached
an intensity level halfway between the core intensity and the wing intensity at a defined wavelength offset from the local core position.
The wing offsets used are $\pm$ 0.1 nm~for the H$\alpha$ line
and $\pm$ 0.06 nm~for the \ion{Ca}{2} 854.2~nm data. The resulting
H$\alpha$ line width maps are shown 
in the middle column in Figure~\ref{fig:IBIS_FOV}.
These line widths are thought to be related to
the temperature of the emitting plasma under
chromospheric conditions
\citep{2009A&A...503..577C, 2012ApJ...749..136L, 2019ApJ...881...99M}.

We measured several different velocity signatures from our 
spectra.
After sampling onto an evenly spaced wavelength grid with 0.005 nm sampling, we fitted the line
minimum
position and intensity with a parabola, in order to
determine the Doppler 
velocity, which is related to the velocity at the
$\tau$=1 for the line core in the atmosphere. 
The fitting was done on an interval of $\pm0.02$ nm and $\pm0.015$ nm for around the minimum position for for H$\alpha$ and \ion{Ca}{2} 854.2, respectively. This corresponds to approximately four (six) of the originally sampled points for H$\alpha$ (\ion{Ca}{2} 854.2), indicated in Figure~\ref{fig:Fast_scan_spec_profiles} as the blue region.

We also calculated the center of gravity (COG)
velocity of over $\pm$~0.12 nm wavelength region for 
the H$\alpha$ line and over a
$\pm$ 0.105~nm for the \ion{Ca}{2} 854.2~nm line 
This velocity measure takes 
into account the whole line profile
and might carry some information about 
the photospheric velocity field in the case of H$\alpha$
\citep{2004ApJ...603L.129S}. 
Finally, we used the same bisector calculation described above to determine a bisector shift
for both lines at 50\% level. 

\subsection{ALMA observations}
\label{subsec:ALMA_observations}

\begin{figure*}
	\includegraphics[width=\textwidth]{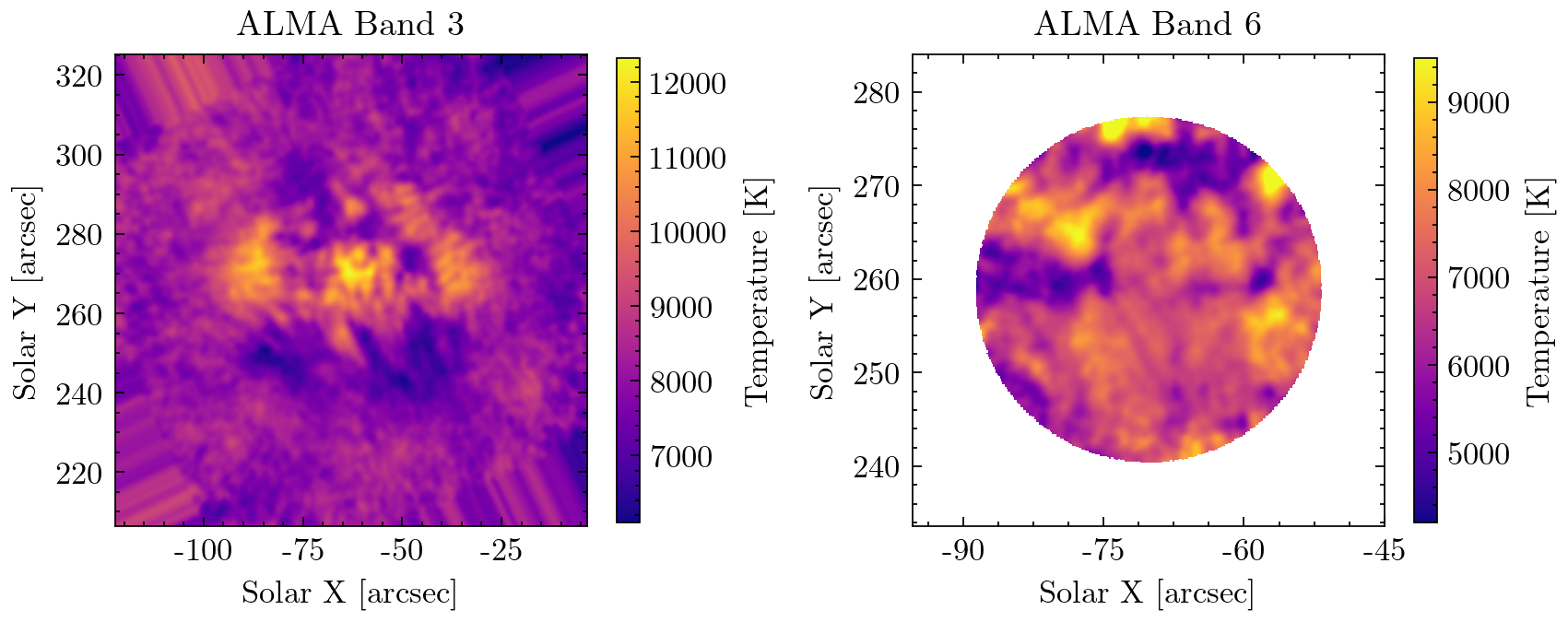}
	\caption{Average ALMA brightness
		temperature maps for Band 3 (left panel) and
		 Band 6 (right panel) for FOV 2. The Band 3 image is the temporal
		 average of observing block 2
	     (taken between 17:31--17:41 UT). The Band 6 image is the
	     average of observing block 2 (taken between 16:03-16:11 UT).
	     The central regions of these observations have the highest
	     sensitivity, where a circular mask was applied to the Band 6 data to 
	     emphasize this region.}
	\label{fig:ALMA_FOV}
\end{figure*}

ALMA became available for
solar observations in Cycle 4 (2017) after
extensive testing
\citep{2015ASPC..499..347P}. The continuum wavelength bands available for solar
observations were Band 3 (3 mm/100 GHz) and Band 6 (1.25 mm/240 GHz)
which are expected to sample the high and
middle chromosphere~\citep{2016SSRv..200....1W}. The
continuum radiation in these wavelengths forms
in the chromosphere under Local Thermodynamic Equilibrium 
(LTE) conditions,
as the main source of opacity is due to free-free processes which 
makes the source function to be locally determined by the plasma temperature.
The ALMA intensity can be
interpreted as brightness temperature under the Rayleigh-Jeans
limit~\citep{2016SSRv..200....1W}.
Hence, ALMA is a valuable tool to study the thermal
structure of the chromosphere 
\citep[some recent results include][]{2017ApJ...841L...5S,2019ApJ...877L..26L, 2020A&A...643A..41D}.

ALMA is an excellent instrument for studying high-frequency waves in the solar atmosphere
due to its fast sampling cadence of 2 seconds,
spatial resolution better than 2$"$ (depending on
the array configuration), 
and direct sensitivity to electron
temperature. However, it is important to remember that the
opacity scale of the ALMA radiation is determined through the local
electron density (and the ionization state of the plasma), which is
thought to be far from thermodynamical
equilibrium~\citep{2002ApJ...572..626C}. 
The time-varying opacity can complicate the interpretation 
of the time-series of temperature brightness measurements 
\citep{2019ApJ...881...99M}.

The ALMA observations discussed in this paper
were obtained in configuration C40-3 and their reduction is described
in \cite{2019ApJ...881...99M}. 
The millimeter observations are obtained in approximately 10-minute blocks interspersed with 
several minutes of off-target calibrations.
%
For FOV 2, we analyzed
temporal blocks 1 (17:19--17:29 UT) and 2 (17:31--17:41 UT) from
Band 3 \citep[also used in][]{2019ApJ...881...99M}
and blocks 1 (15:53--16:01 UT) and 2 (16:03--16:13 UT)
from Band 6. 
An ALMA
data summary is presented in Figure~\ref{fig:ALMA_FOV}.
We chose these particular ALMA observing blocks because they 
are the closest in time to
our 
high-cadence IBIS observations of FOV 2 listed in Table~\ref{table:Observations}.
The relative positions of the ALMA observing fields to the IBIS
observation regions are
shown in Figure~\ref{fig:context_1} as the colored
circles. The useful regions of the FOVs
of the ALMA data are about the
size of the respective circles 
in Figure~\ref{fig:context_1}:
60{\arcsec} for ALMA Band 3 and 
20{\arcsec} for Band 6.

\subsection{Solar feature classification}
\label{subsec:solar_feature_classification}

To study the typical wave 
characteristics in different regions of the
solar atmosphere, we partitioned the solar surface in the field
of view into five different classes
of features that 
represent general chromospheric structures: 
penumbra, internetwork, fibrils, network and plage.
We distinguished the regions 
following a methodology based on the properties of the  
H$\alpha$ and \ion{Ca}{2} 854.2 nm spectral lines.

For FOV 2, we first categorized the penumbral region by its proximity
to the leading sunspot in NOAA AR 12651 and its low H$\alpha$ 
line width (smaller than 0.1 nm)
and low Doppler velocity fluctuations relative to the
rest of the field.
Secondly, we distinguished between the plage and the network regions, which 
are brightest in FOV 2. We can clearly distinguish
between the two by the amount of 
magnetic flux and the intensity in the \ion{Ca}{2} 854.2~nm line core intensity.
We assign a pixel to be a plage region if 
the photospheric magnetic field strength (from HMI) is above 1200 G
and the \ion{Ca}{2} 854.2~nm line core intensity is 
more than 60\% above 
the mean intensity of the whole FOV. 
Pixels were categorized as network if their magnetic 
field was above 250 G and their \ion{Ca}{2} line core intensity
is below the 60\% 
intensity threshold. 
We labeled the regions with lowest 
intensity in the core of \ion{Ca}{2}
854.2 nm and with the lowest H$\alpha$ line core width 
(less than 0.105 nm) as internetwork or fibrils.
We distinguished between the fibrils and the 
internetwork regions by the ratio of the relative power in the 3 min to 
5 min power in the Doppler velocity power spectra as suggested by
\cite{2007A&A...461L...1V}, where the 
internetwork has a ratio greater than one and the fibrils
have a ratio less than one between those two 
frequency windows.

\begin{figure}
	\includegraphics*[width=.475\textwidth]{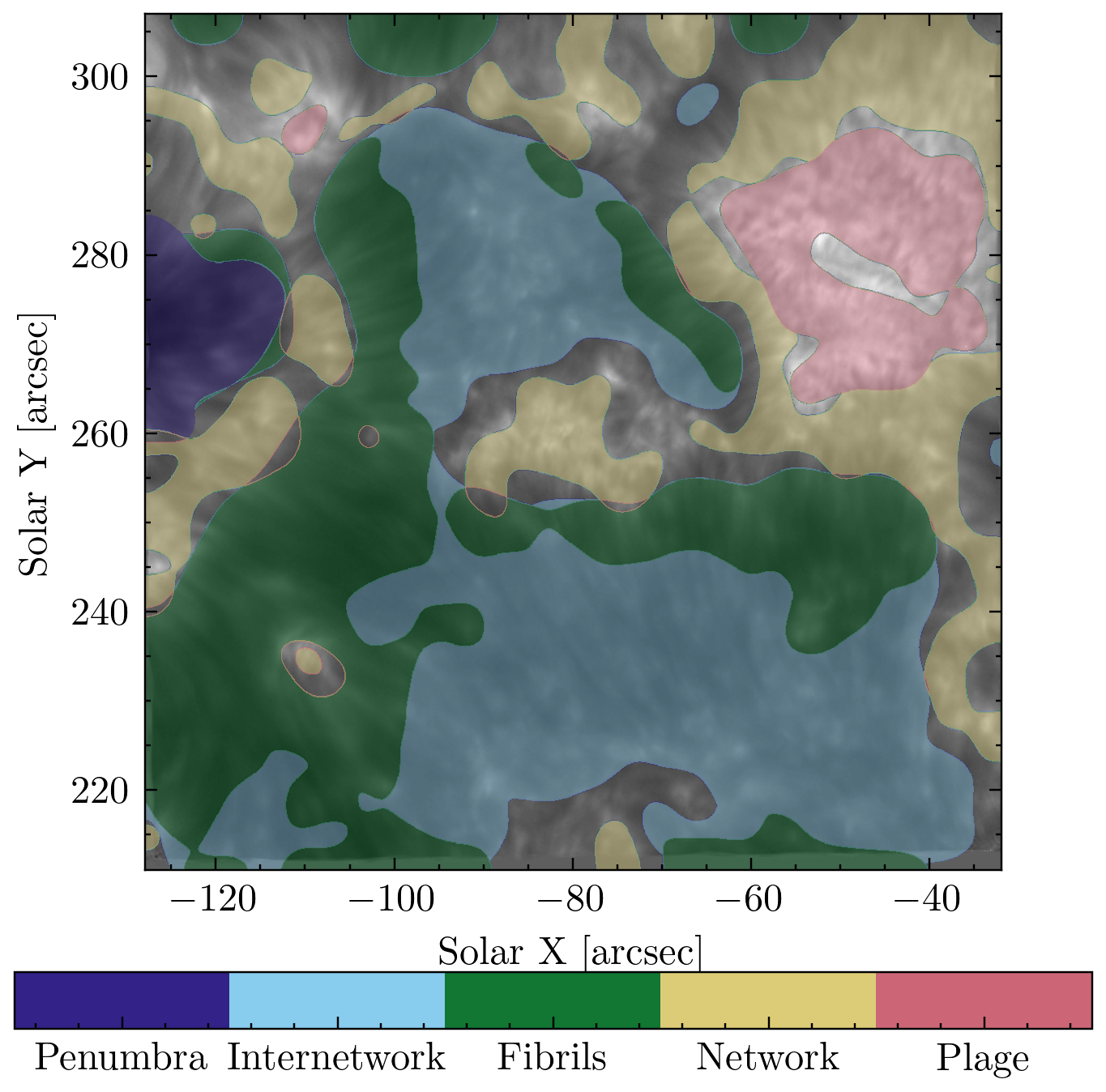}
	\caption{Mask for FOV 2 for the different regions 
	    of the solar surface (shaded in 
		the corresponding color) overlaid over the \ion{Ca}{2} line
		core intensity. 
		}
	\label{fig:masks_regions_of_sun}
\end{figure}

Each of these criteria resulted in binary masks, which we blurred with 
a Gaussian filter (with a
standard deviation of 40 pixels) to smooth the boundaries of the 
regions and avoid 
holes. In cases where the smoothing caused masks for separate classes
to overlap, we chose the darker of the two classes to define that 
pixel. The blurring insured, for example, that 
isolated magnetic elements in the 
internetwork were smoothed out and the classified regions 
were largely contiguous.
The resulting mask for FOV 2 which we employed throughout
the rest of the paper to distinguish the different regions 
of the solar
surface is shown in Figure~\ref{fig:masks_regions_of_sun}. 

\begin{figure}
	\includegraphics*[width=0.475\textwidth]{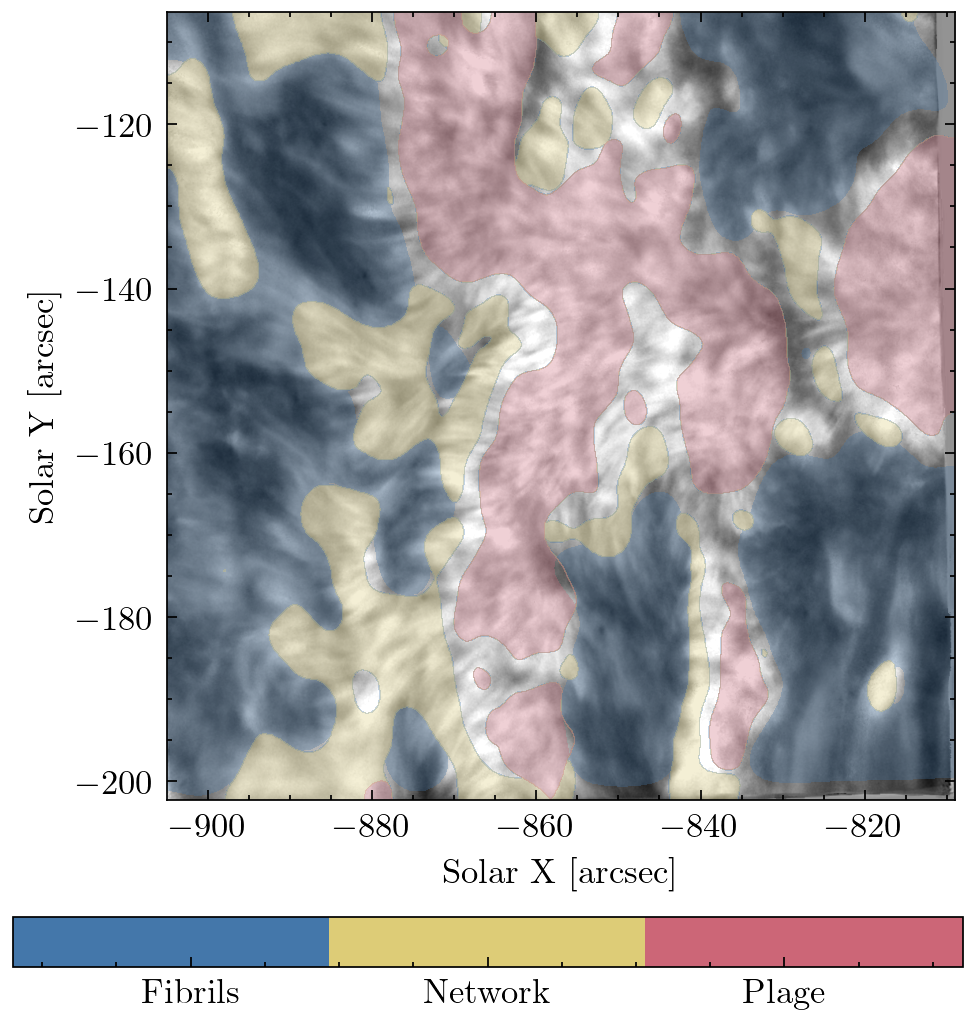}
	\caption{Mask of the different solar regions in FOV 1
	using the approach 
		described in Section~\ref{subsec:solar_feature_classification}. 
		No distinction between 
	    internetwork and fibrils was made for this FOV due to the large viewing angle and preponderance of magnetic flux.}
	\label{fig:FOV_1_mask}
\end{figure}

We use a similar masking algorithm for the FOV 1 observations, 
but using only
three types of solar
features because there is no penumbra in this target and we cannot easily distinguish between 
the internetwork and the fibrils at such highly inclined 
projections. We take into account in our analysis that FOV 1 is mostly
a plage region and hence label all of the ``quieter'' regions as
fibrils. 
We increased the H$\alpha$ core width cutoff for fibrils
to 0.12~nm, as the average H$\alpha$ line
width increases closer 
to the solar limb. We 
distinguished between 
plage and network by the photospheric magnetic field
strength -- if the magnetic field was above 1200 G (as for FOV 2)
then we classified the pixel as plage, otherwise it was classified 
as network. We also used a \ion{Ca}{2} 854.2 nm line core
intensity threshold 
that was twice the average intensity of the 
whole FOV
for distinguishing between plage and 
network
The resulting mask for FOV 1 is presented 
in Figure~\ref{fig:FOV_1_mask}.

\section{Observed high-frequency power in the spectral 
	diagnostics} 
\label{sec:High_frequency_power}

In this Section we present the high-frequency observations derived 
from the Power Spectrum Density (PSD) 
of the observed chromospheric diagnostics (intensities, velocities,
 etc.) described in Section~\ref{sec:Observations}.
The PSDs are calculated as the absolute value of the 
Fourier amplitudes of the observed signals,
where we subtracted the mean
of the observed time series and then 
apodized them with a Hamming window with 3\% of the
total length of the observations on each side.

\begin{figure*}
\includegraphics[width=.9\textwidth]{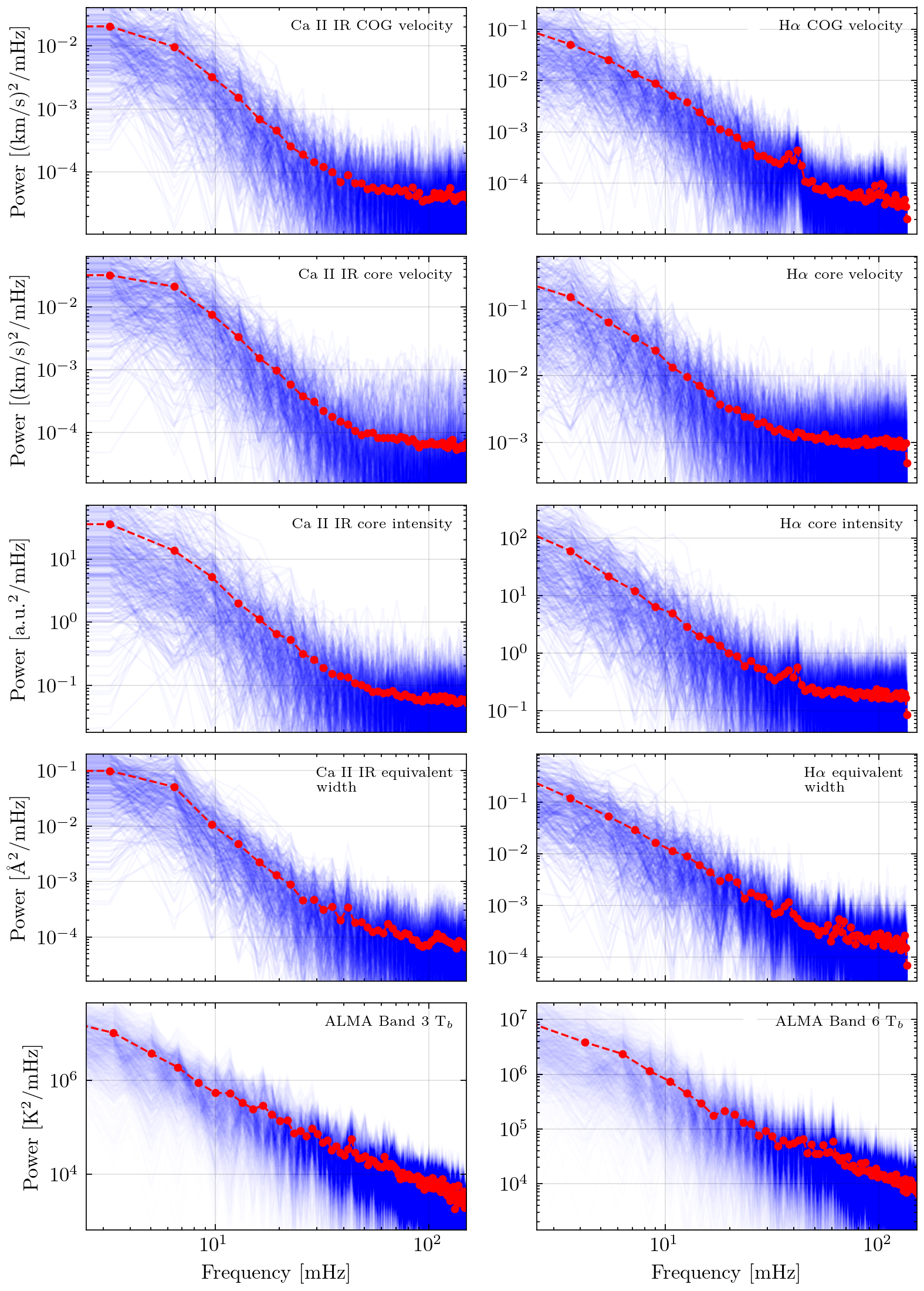}
\caption{Power spectra of different spectral chromospheric
 diagnostics. In the first four
rows, the left columns correspond to spectral properties derived from the 
\ion{Ca}{2} 854.2 nm line (\ion{Ca}{2} IR) and the right column from the H$\alpha$ line
observations. The chromospheric diagnostics
which each row is derived from are: \emph{First row:} COG 
velocity; \emph{Second row:} line center velocity; \emph{Third row:} 
intensity of the line core (where a.u. stands for arbitrary units);
\emph{Fourth row:} equivalent width. 
The last row corresponds to PSDs derived from ALMA
Band 3 (\emph{left row}) and Band 6 (\emph{right row}) intensity observations.
The blue lines represent individual power spectra and the
red dotted lines show the mean in each frequency bin.}
\label{fig:Power_spectra_overview}
\end{figure*}

\subsection{Observed high-frequency power in chromospheric spectral
diagnostics}
\label{subsec:spectral_obs_IBIS}

The top four rows of Figure~\ref{fig:Power_spectra_overview}
present a summary of the power spectra of the chromospheric spectral 
diagnostics derived from H$\alpha$ (data series starting at 15:18:57 UT) and 
\ion{Ca}{2} 854.2 nm line (data series starting at 15:13:21 UT) IBIS observations of FOV 2.
The pixels analyzed for this plot were taken from the 
central 12\arcsec$\times$ 12{\arcsec} regions of FOV 2 
(the green rectangle in Figure~\ref{fig:IBIS_FOV})
to ensure that the same solar region is observed 
with IBIS and ALMA. Furthermore, the selected area
(green square) spans only a network region, which simplifies the
analysis to a single type of source region.
The red dotted line 
is the mean of the distribution at each frequency and the lighter blue lines are 
individual power spectra of each pixel.
The frequency resolution of the power spectra is on the order of
$\delta \nu \approx$ 2 - 3 mHz as the length of our data series 
is about 10 minutes and the cadence is on the order of 3.3 seconds
(see Table~\ref{table:Observations} for the details of 
the observations).
We do not clearly observe the low frequency power
around the five-minute 
oscillation window with high resolution due to the 
short temporal extent of these fast scans.

The power spectra show a strikingly similar 
power law above 7 mHz in all observed spectral diagnostics, limited 
at the higher frequencies by white-noise floor. 
The level of the white noise 
varies for the different spectral diagnostics. 
Our results
extend previous work by \cite{2008ApJ...683L.207R} to higher frequencies where
they agree in the low frequency
part (7--20 mHz) of the power spectrum.
The white noise which 
dominates the high frequency signal is likely due primarily to the photon noise from the measured signal.
In Appendix~\ref{App:Noise_floor} we describe a detailed estimation
of the effect from photon noise on the 
white-noise floor of the line core velocity measurements of the 
\ion{Ca}{2} 854.2~nm data. 
This additional contribution to the PSD is important to characterize in order to
properly quantify only the solar contribution to the integrated power. 

\begin{figure}
	\includegraphics*[width=.45\textwidth]{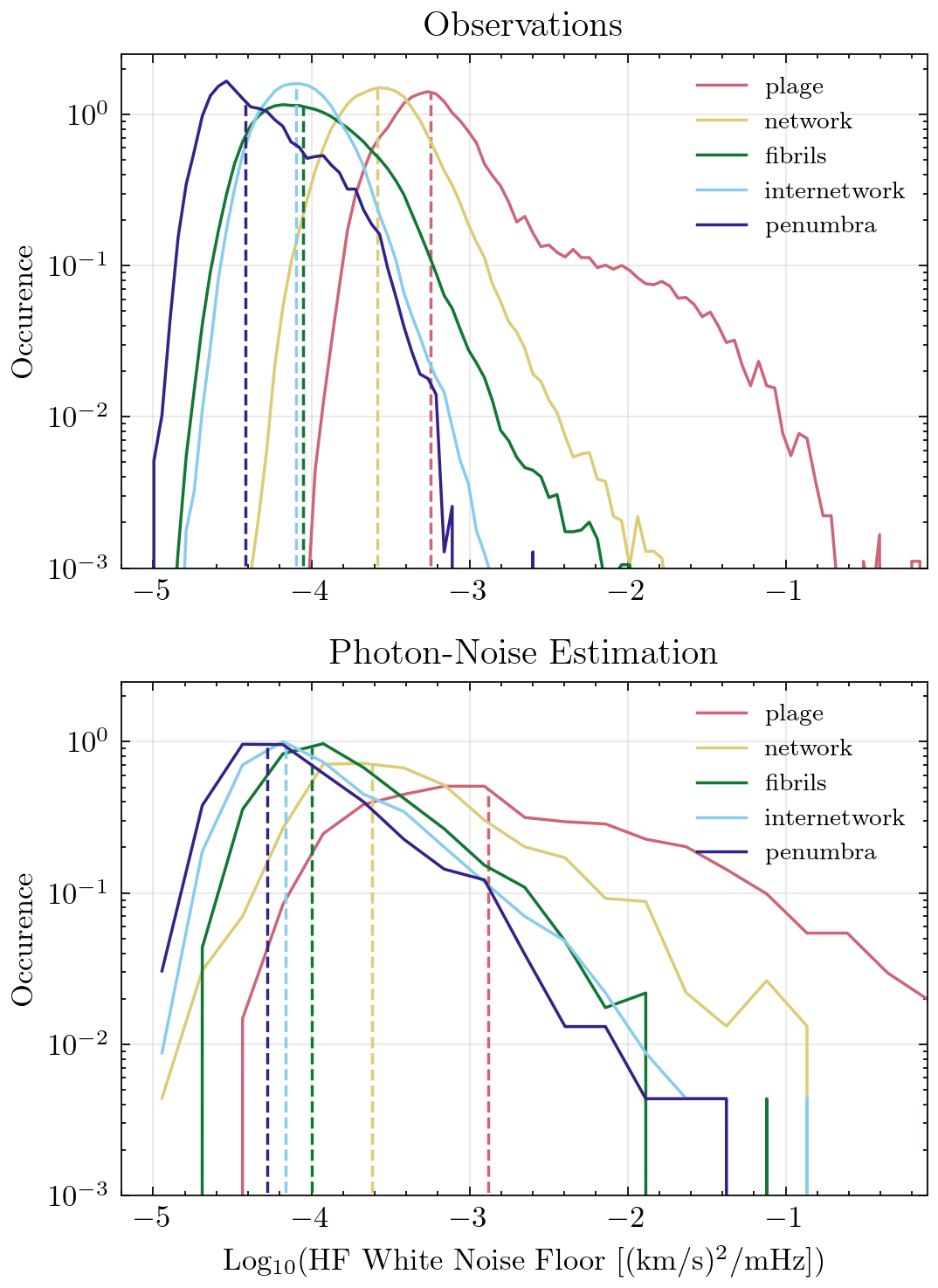}
	\caption{
		Histogram of the high frequency white noise 
		level for different regions of the solar 
		surface. 
		 \textbf{Top panel:}
		White noise limit 
		in the IBIS \ion{Ca}{2} 854.2 nm
		data measured as the median of the last 25 
		frequency bins in the PSDs of the individual pixels.
		\textbf{Bottom panel:} Estimate of the white
		noise floor from photon noise based on the 
		method presented in Appendix~\ref{App:Noise_floor}.
		The solar surface regions were distinguished as 
		shown in Figure~\ref{fig:masks_regions_of_sun}.
		The median of each distribution is presented as the dashed 
		vertical line.}
	\label{fig:HF_noise_levels}
\end{figure}

The observed and simulated white-noise level distributions
are shown in the Figure~\ref{fig:HF_noise_levels} and summarized in 
Table~\ref{table:ROS_flux_summary}.
In both cases, we see that the profiles with higher 
\ion{Ca}{2} 854.2 nm line core intensities (plage, network) tend to have higher noise levels compared to deeper profiles (with more pronounced core minima). 
The distributions of the simulated noise levels have somewhat broader
tails, indicating a possible overestimation of the noise in our model
in extreme cases. However, the median values (dashed vertical lines) 
of the simulated distributions are all in good agreement with the
observed ones.
Hence, we believe that we can well characterize the noise floor separately for different classes of solar structures in our dataset. 

We employ the same data analysis approach 
for the brightness temperatures from
the ALMA Bands 3 and 6 data. 
The PSD of the brightness temperature 
from the same central 12\arcsec$\times$ 12{\arcsec}
of the ALMA FOVs are
presented in the last row of Figure~\ref{fig:Power_spectra_overview}.
The red lines again indicate the average of the 
distribution and the light blue lines are the power
spectra of individual pixels.

The power law behavior of the ALMA brightness temperature PSD
(seen before in~\cite{2020A&A...638A..62N})
extends to the white-noise floor 
at about 100 mHz in the Band 3 data.
In the Band 6 observations we do not see clearly the white-noise floor.
We do not observe the 3 minute (5 mHz) oscillations clearly, like 
in~\cite{2020A&A...634A..86P}. This might be due to 
presence of magnetic elements in the observed region as suggested 
by~\cite{2021RSPTA.37900174J},
or due to the limited frequency
resolution of our data of about 2 mHz.
However, the observed region 
is mostly covered under the magnetic canopy where
we notice the bright fibrils (in ALMA wavelengths) 
dominating our field of view
(see Figures~\ref{fig:context_1} and \ref{fig:IBIS_FOV}). 

\subsection{Properties of the observed power spectra}
\label{subsec:power_spectra_properties}

\begin{figure}
	\includegraphics*[width=.45\textwidth]{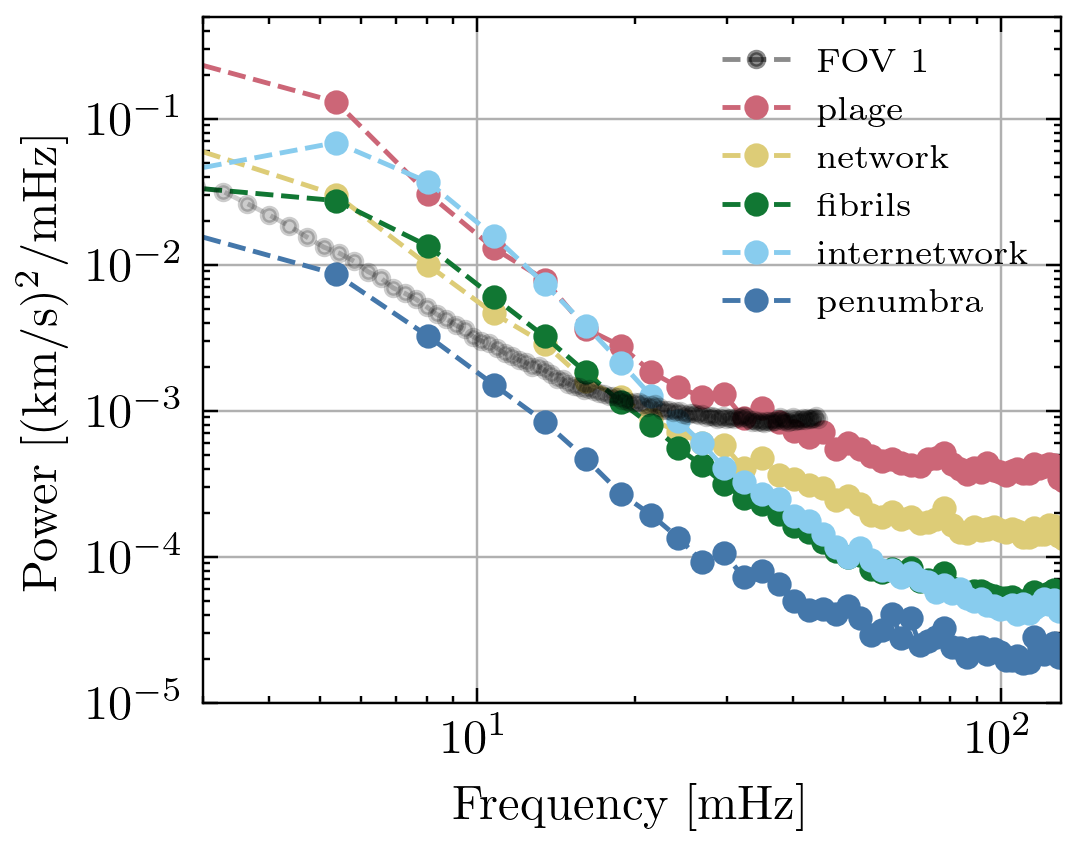}
	\caption{Averaged power spectra of the Doppler line core velocity
		from \ion{Ca}{2} 854.2 nm line for different solar   
		regions in FOV 2. The regions are outlined
		in Figure~\ref{fig:masks_regions_of_sun}. The median
		FOV 1 \ion{Ca}{2} Doppler velocity PSD 
	    is shown as the semitransparent black dotted line.}
	\label{fig:average_PSD}
\end{figure}

We concentrate on the \ion{Ca}{2} 854.2~nm velocity power
spectra as they are the most reliable chromospheric
velocity diagnostic we have, as the 
H$\alpha$ line synthesis results depends strongly on the full 3D radiative transfer solution~\citep{2012ApJ...749..136L}.
Using the feature classifications
derived in Section 
\ref{subsec:solar_feature_classification}
and shown in Figure~\ref{fig:masks_regions_of_sun},
we calculated the average \ion{Ca}{2} 854.2 nm 
line core Doppler velocity PSD 
for every type of region of the solar surface described above. The 
average PSDs are presented in Figure~\ref{fig:average_PSD}. 
The plage and internetwork have higher amounts of 
velocity oscillation power 
compared to the network and the quieter regions (fibrils 
and penumbra). However, the plage 
has a significantly higher  
white-noise floor (comparable to the network one) than the one 
seen in the internetwork. 
The internetwork has high amount of wave flux in the 5-20 mHz interval
but a lower white-noise floor -- similar to the one in the quieter regions (fibrils, penumbra).

\begin{table*}
	\begin{center}
		\begin{tabular}{| c | c | c | c | c|}
		    \hline
			Solar region & Oscillatory power & Raw/Corrected& Log$_{10}$(Noise floor) & Acoustic flux \\
			& [(km/s)$^2$] &Power law slope& [(km/s)$^2$/mHz] & [W/m$^2$] \\ \hline
			Penumbra & 0.034$^{+0.097}_{-0.024}$ & $-2.47^{+1.33}_{-1.42}$ / $-0.86^{+1.07}_{-1.46}$
			& $-4.41^{+0.54}_{-0.28}$ &  
			24$^{+16}_{-45}$ \\ \hline
			Internetwork & 0.47$^{+0.59}_{-.26}$ & $-3.42^{+1.63}_{-1.21}$ / $-1.82^{+1.31}_{-1.22}$ &
			$-4.12^{+0.35}_{-0.29}$ & 203$_{-119}^{+228}$ \\ \hline
			Fibrils & 0.17$^{+0.30}_{-0.12}$ &$-2.92^{+1.46}_{-1.35}$ / $-1.32^{+1.22}_{-1.34}$ &
			$-4.04^{+0.47}_{-0.39}$ 
			& 102$^{+138}_{-59}$ \\ \hline
			Network & 0.16$^{+0.36}_{-0.11}$ &$-2.14^{+1.33}_{-1.30}$ / $-0.62^{+1.00}_{-1.32}$ &
			$-3.58^{+0.36}_{-0.32}$
			 & 109$_{-71}^{+193}$ \\ \hline
			Plage & 0.48$^{+1.46}_{-0.38}$ & $-2.39^{+1.07}_{-1.11}$ / $-0.79^{+0.89}_{-1.14}$ &
			$-3.25^{+0.76}_{-0.34}$ 
			&  256$^{+847}_{-166}$ \\ 
			\hline
		\end{tabular}
		\caption{Summary of the power law properties and the
			wave flux observed in the
			different regions on the solar surface for FOV 2 \ion{Ca}{2} 
			854.2 nm data (15:39:54 UT).}
		\label{table:ROS_flux_summary}
	\end{center}
\end{table*}

\begin{figure*}
	\includegraphics*[width=\textwidth]{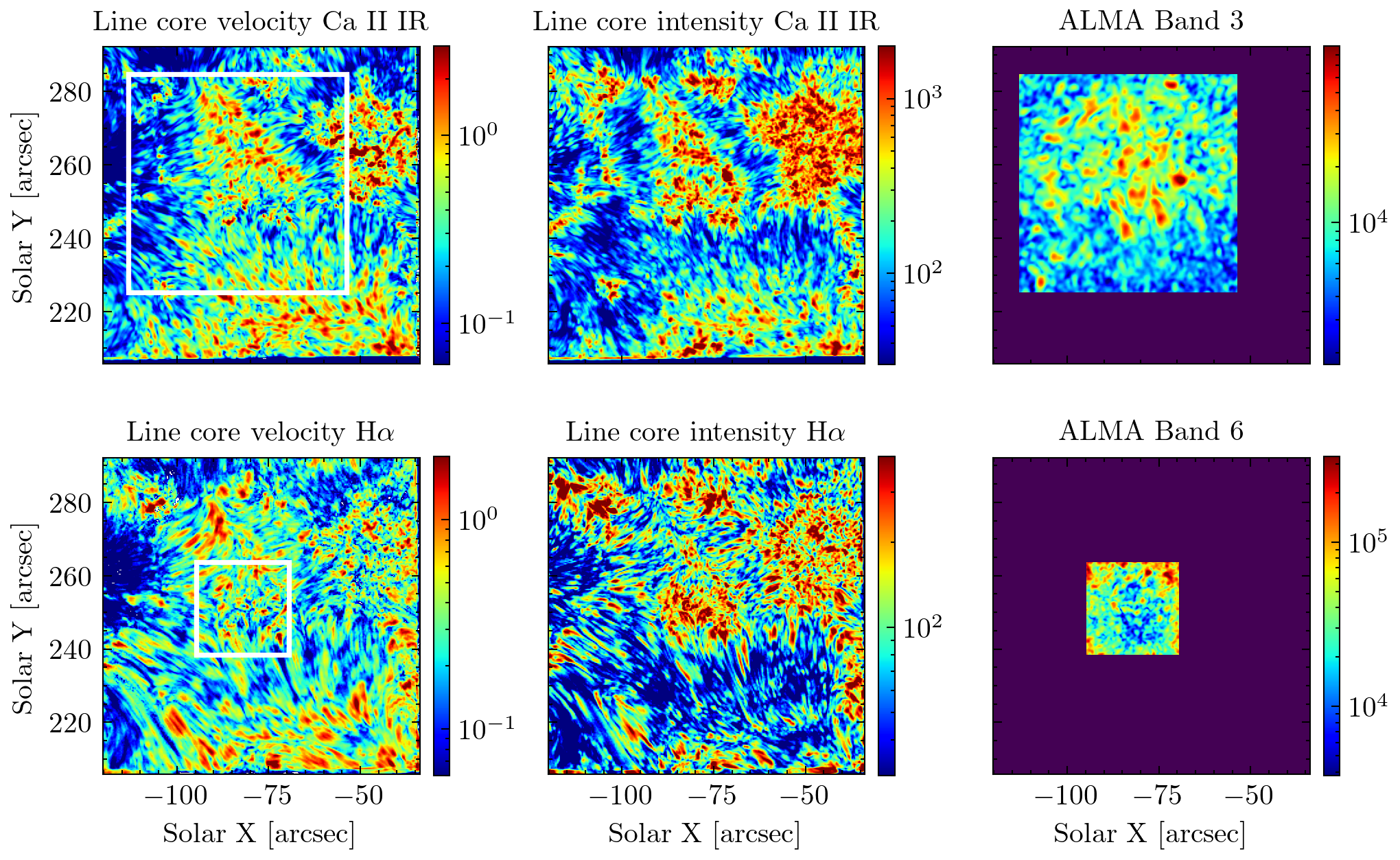}
	\caption{Total oscillatory power between 5 mHz and 50 mHz 
	    for different chromospheric diagnostics in FOV 2. The panels
	    show the following quantities:
		\textbf{Top row}: \emph{Left}: \ion{Ca}{2} 854.2 nm line 
		center velocity power [(km/s)$^2$];
		\emph{Center:} \ion{Ca}{2} 854.2 nm line 
		center intensity power [counts$^2$];
		 \emph{Right:} ALMA Band 3 brightness temperature [K$^2$]; 
		 \textbf{Bottom row:} \emph{Left}: H$\alpha$ 656.3 nm line 
		center velocity power [(km/s)$^2$];
		\emph{Center:} H$\alpha$ 656.3 nm line
		center intensity power [counts$^2$];
		 \emph{Right:} ALMA Band 6 brightness temperature [K$^2$]. The
		 white panels in the left column show the FOV 
		 of the ALMA Band 3 (top row) and Band 6 (bottom row). 
	}
	\label{fig:power_maps}
\end{figure*}

Figure~\ref{fig:power_maps} shows the spatial maps for FOV 2 of the amount of oscillatory power at each pixel integrated between 
5 and 50 mHz in the line core intensity and 
Doppler velocity measured from the
\ion{Ca}{2} 854.2 nm and H$\alpha$ lines as 
well the oscillatory power derived from the
ALMA brightness temperatures.
Throughout the paper when we refer to oscillatory
power we mean the integrated power spectral density (PSD) in the
specified frequency range. When estimating PSDs, we always
subtract a white noise estimate based on the mean of the
last 25 (high) frequency bins.
We see that
properties of similar diagnostics (velocities, intensities and widths) derived from 
different spectral lines have similar distributions.
The ALMA Band 3 temperature fluctuations map
(top right)
correlates with the spatial distribution 
of the PSDs in
the optical diagnostics, but the Band 6 one (bottom
right) does not appear to resemble 
its optical counterparts. This might be due to time-varying changes in the height of formation of the 
ALMA intensities as well as the lower angular resolution of the
ALMA observations or the very limited FOV of the ALMA Band 6 observations.

\begin{figure}
	\includegraphics*[width=0.475\textwidth]{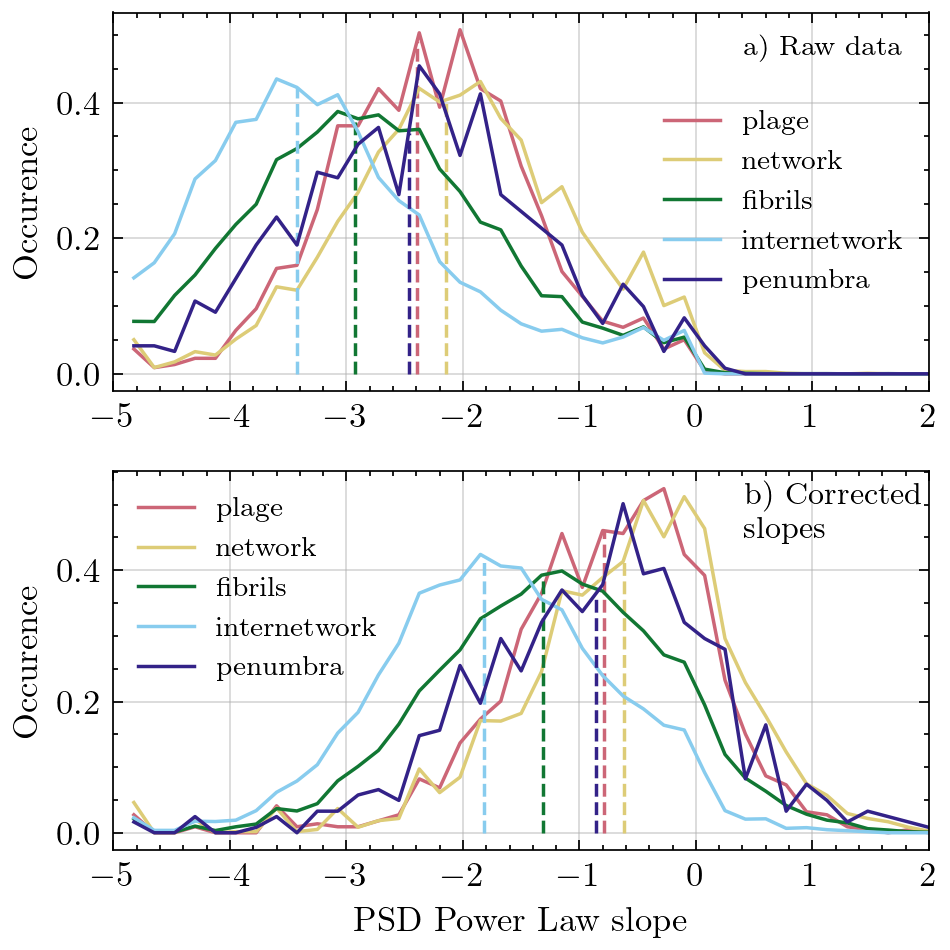}
	\caption{Histogram of the slopes of the power law fit to the 
	         PSD of the Doppler velocity
		     of \ion{Ca}{2} 854.2 nm line for \textbf{Panel a)}: Raw data; 
		     \textbf{Panel b)}: Corrected data (see Section~\ref{subsec:power_spectra_properties}).
		     The fit was made between 10 mHz and 40 mHz.
		     The dashed lines show the median of the distributions.}
	\label{fig:PSD_slope_histogram}
\end{figure}

We calculated the power law slope describing the observed power spectra 
with least-$\chi^2$ fitting
between 10 and 40 mHz. 
Using the same solar feature mask,
we calculated the distributions 
of the slopes in the \ion{Ca}{2} 854.2 nm line core 
velocity in the different
regions of the solar surface and the results are presented in
Figure~\ref{fig:PSD_slope_histogram}. The distributions of the slopes from the observed 
(raw) data are presented in Panel a) of Figure~\ref{fig:PSD_slope_histogram}.
We see that plage and network regions have shallower
slopes than the fibrils, internetwork and penumbral 
regions.
The slope
distributions are overlapping and 
form a continuous transition between all of 
%
the different types of solar features, but their progression does 
follow the trend of mean H$\alpha$ line-widths for those solar 
features (excluding the penumbra).
The quantitative comparison of the power slope
distributions is 
listed in Table~\ref{table:ROS_flux_summary}, along with some 
other quantities derived from the observed power laws.
The quoted regions of uncertainty are 
the 10th/90th percentiles of the
cumulative distributions. We note that the power-law slopes we find
for the network, fibril, and internetwork regions are very comparable 
to those found by \cite{2008ApJ...683L.207R} for the 5-20 mHz interval.  

To infer the true slope of the vertical velocity field in the solar atmosphere,
we applied the velocity attenuation coefficient 
described in Section~\ref{subsec:RADYN_T_CaII}. The resulting corrected velocity power law slope  
distributions are presented in Panel b) of Figure~\ref{fig:PSD_slope_histogram} and their medians are 
summarized in Table~\ref{table:ROS_flux_summary}. The relative order of the solar features is 
preserved, but the resulting distributions of slopes are
shallower as the compensation for the  wave 
attenuation makes the power laws less steep. The resulting 
values of the power law slopes of the corrected data are roughly between $-2$ and 0. 
These values are shallower
than the ones expected from the 
Lighthill-Stein turbulence theory~\citep{1996A&A...315..212U}, 
which predicts power law slopes between $-3.5$ and $-3$. 
Eulerian-based treatments of turbulence in the solar 
atmosphere give slopes between $-2.4$ and $-1.3$, which 
are closer to our observed values~\citep{2002ApJ...572..674R}. Hence, our 
observations favor the Eulerian approach in treating the turbulent 
time correlations in the solar atmosphere. 
However, the penumbra, the network, and the plage regions which exhibit higher 
power law slopes than either theory predicts and require further investigation, which 
we leave for a following study. 

\begin{figure*}
	\includegraphics*[width=\textwidth]{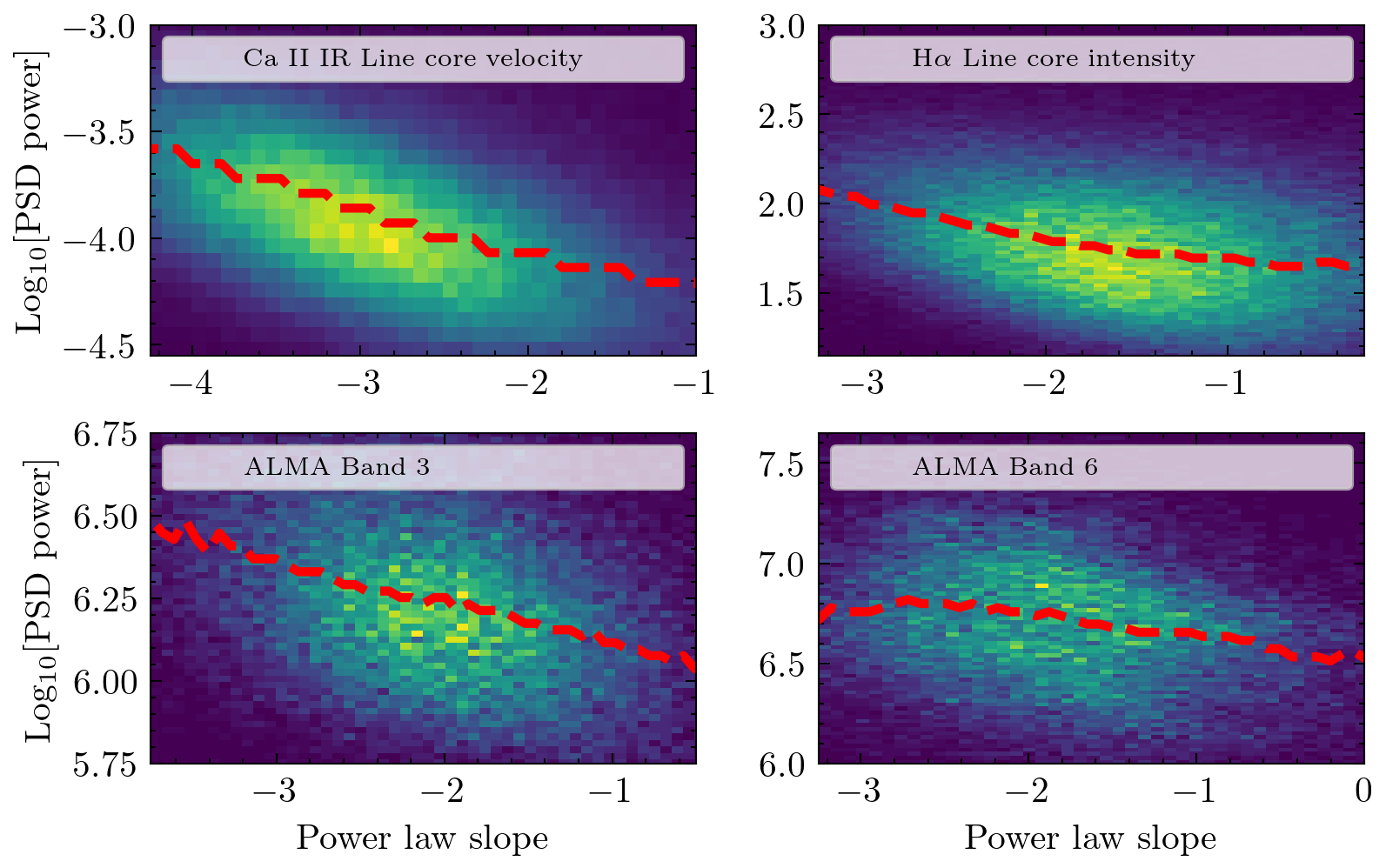}
	\caption{2D histograms of the total oscillatory power in the
	  respective diagnostic against the slope of the power law fit in the
	  frequency range of 8 to 35 mHz. 
		\textbf{Top}: \emph{Left:} \ion{Ca}{2} 854.2 nm line core
		velocity. \emph{Right:}	H$\alpha$ 656.3 nm line core intensity;
		\textbf{Bottom}: \emph{Left:} ALMA Band 3 brightness temperature; \emph{Right:} ALMA Band 6 brightness temperature.
		The red line represents the average of each column of the histograms.}
	\label{fig:PSD_vs_slope}
\end{figure*}

 Figure~\ref{fig:PSD_vs_slope} 
shows the correlation between the uncorrected power-law slopes and the 
total amount of 
power in each spectral diagnostic, with pixels with higher integrated power having steeper slopes. 
Since the PSDs are steeper, the higher power is a consequence
of those regions tending to have higher overall oscillatory 
power in the lower 
portion (5-10 mHz) of the frequency interval.

\begin{figure}
	\includegraphics[width=0.475\textwidth]{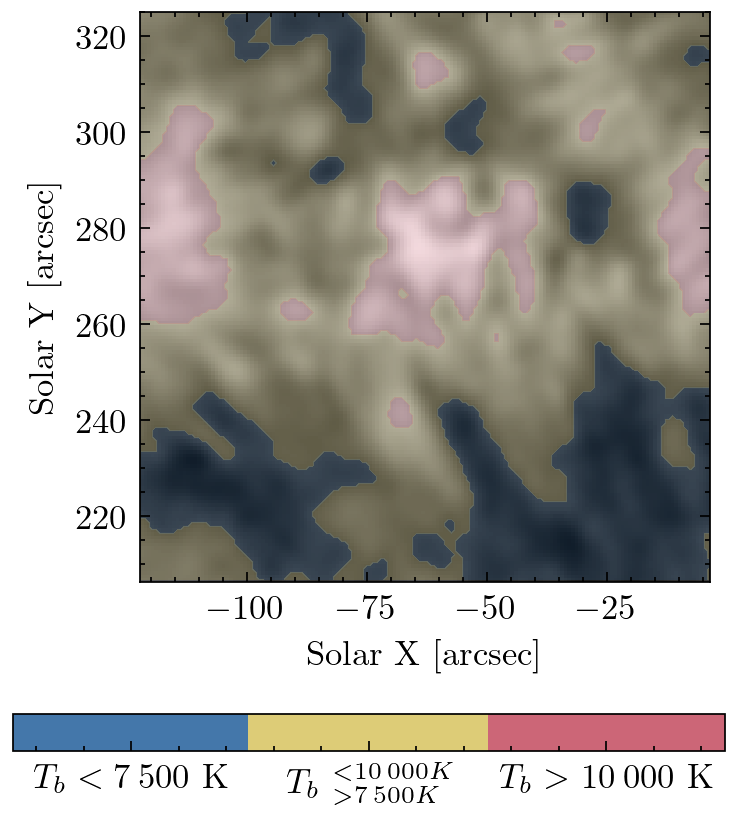}
	\caption{The mask distinguishing the different regions 
		of the solar
		surface in for the ALMA Band 3 FOV. The mask is based on the
		brightness temperature $T_b$ split 
		into three categories -- $T_b < 7\:500$K,
		$7\:500 < T_b < 10\:000$K,
		and $T_b > 10\:000$ K.}  
	   
	    \label{fig:ALMA_mask}
\end{figure}

\begin{figure}
	\includegraphics[width=0.45\textwidth]{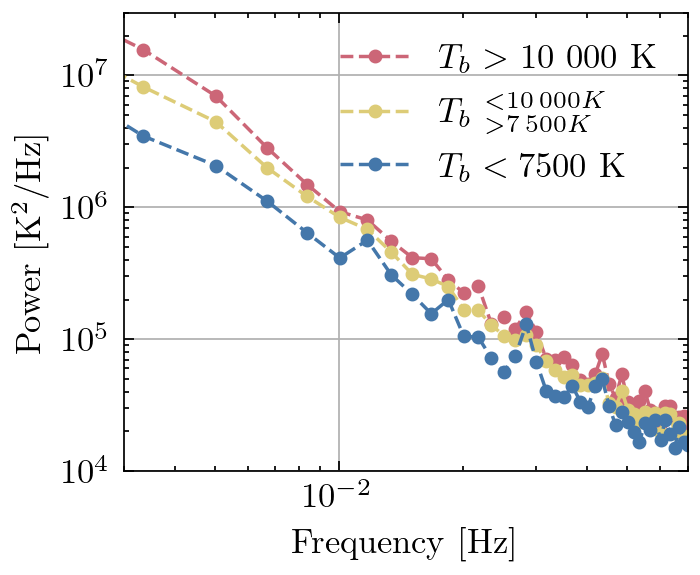}
	\caption{Average ALMA Band 3 brightness temperature $T_b$
		PSDs for different regions of the solar surface 
		as segmented in Figure~\ref{fig:ALMA_mask}.}
	\label{fig:ALMA_psd}
\end{figure}

We partitioned the Band 3 field of view into three regions based on the mean
observed 3 mm brightness temperature: 
$T_b$ $<$ 7500 K; 7500 K $<$ $T_b$ $<$ 10000 K; and 
$T_b$ $>$ 10000 K. The resulting segmentation mask is presented
in Figure~\ref{fig:ALMA_mask}.
%
We used this segmentation approach because the classification 
described in Section \ref{subsec:solar_feature_classification}
did not necessarily correspond
well with the ALMA brightness temperature features. 
%
In  particular, the fibril regions 
exhibit brightness temperatures 
comparable to those seen in the network regions.
%
The average power spectrum of the ALMA Band 
3 brightness temperature for each of these different 
solar regions are presented in
Figure~\ref{fig:ALMA_psd}. The power spectra share
similar slopes, independent of the region, with a value 
of -1.63 $\pm$ .07 in the region of 10 to 50 mHz.
The brighter regions have slightly
higher oscillatory power at frequencies below 10 mHz.
We did a similar analysis of the Band 6 data (using the same temperature-based segmentation 
based on the Band 3 data), but did not find
%
any significant differences in the slopes or total power among
the different (temperature discriminated) regions of that FOV. 
 This might be due to the rather small FOV of Band 6 (20\arcsec) compared to Band 3 (60\arcsec).

\begin{figure}
	\includegraphics*[width=0.45\textwidth]{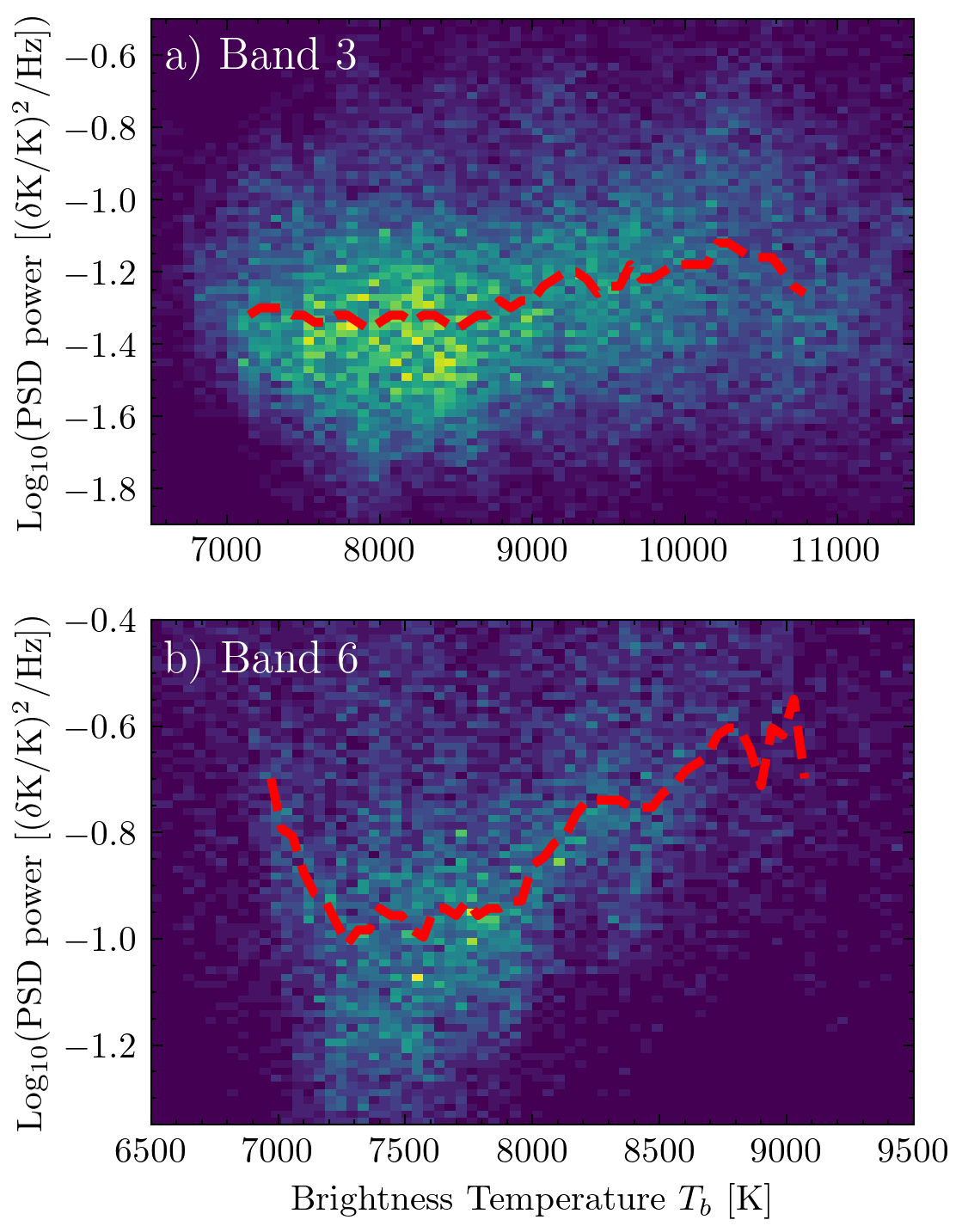}
	\caption{Integrated ALMA brightness temperature 
		 oscillatory power in the frequency range from 5 to 50 mHz
		 against the observed brightness temperature for 
		 \textbf{a)} Band 3 and \textbf{b)} Band 6. 
		The red 
		line shows the running median of the distribution for each 
		column of the histogram.} 		 
	\label{fig:ALMA_PSD_power_vs_Tb}
\end{figure}

%
Figure~\ref{fig:ALMA_PSD_power_vs_Tb} shows the correlation
between brightness temperature and the relative (normalized to the 
mean) brightness temperature
fluctuations in the ALMA data between 5 and 50 mHz.
ALMA Blocks 2 were used for both bands in the figure. In order to 
avoid the attenuation of the sensitivity farther away from
the center of the beam, we used only the central 60{\arcsec}
of the Band 3 FOV and the central 17{\arcsec} for the Band 6
FOV. The red lines show the median 
trend of the histograms. We can see a clear correlation between the
brightness temperature and oscillatory power
in the Band 6 data 
and an almost non existent
one in the Band 3 data. The relative
temperature fluctuation power 
is related to the amount of compressive
wave flux as shown in Section~\ref{subsec:Transmission_coeff_RADYN}.

\subsection{Velocity oscillations at different viewing angles}
\label{subsec:FOV1_power}

\begin{figure}
	\includegraphics*[width=0.475\textwidth]{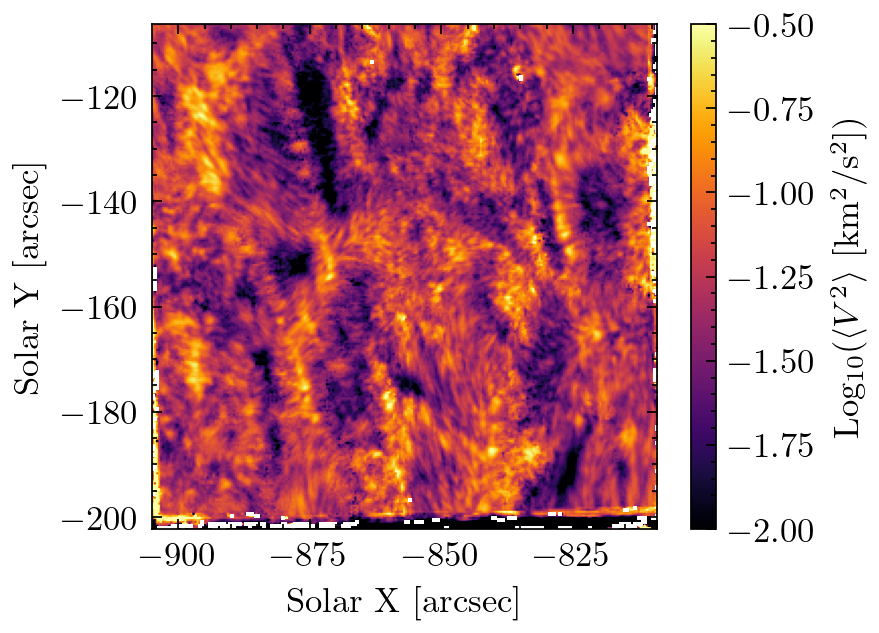}
	\caption{Velocity oscillations map for FOV 1 derived from
		the \ion{Ca}{2} 854.2 nm line core Doppler velocity between 5 and 20 mHz.
		The white-noise floor has been 
		subtracted as described in Section~\ref{subsec:DST_observations}.}
	\label{fig:FOV_1_power}
\end{figure}

\begin{figure}[ht!]
	\includegraphics*[width=0.475\textwidth]{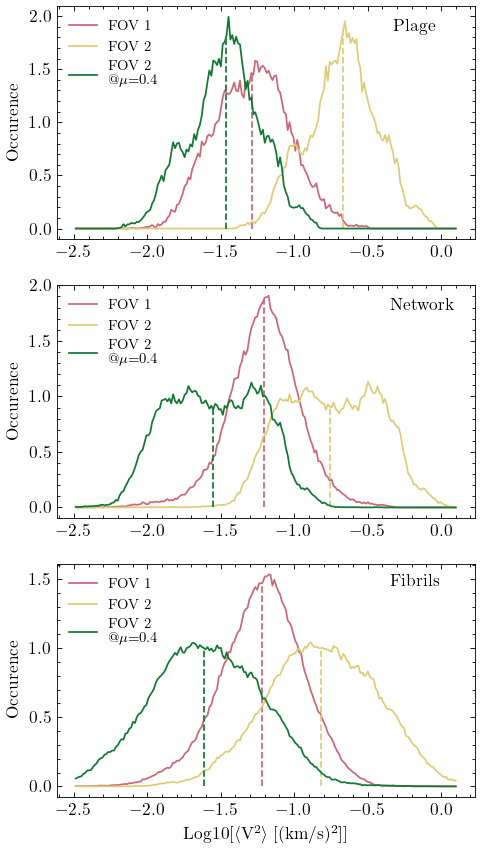}
	\caption{Distributions of the total velocity oscillatory power
		between 5 mHz and 20 mHz in the two 
		fields of view for different solar surface features
		inferred from the \ion{Ca}{2} 854.2~nm line.
		The green curve 
		is the PSD of FOV 2
		if observed at incidence angle $\mu$=0.41 
		(see Section~\ref{subsec:FOV1_power}).
		The vertical dotted lines show the median of the distributions.}
	\label{fig:FOV_power_histograms}
\end{figure}

Observing the solar atmosphere at different inclinations
provides a way to disentangle the longitudinal 
from the transverse velocity oscillations, as these two components are differently projected into line-of-sight Doppler shifts. 
This allows us
to statistically disambiguate between the 
transverse and the longitudinal oscillations if we
observe similar solar regions at (a minimum)
two different viewing angles.
FOV 1 was observed at an incidence angle of 66 degrees or 
$\mu=0.41$, close to the east solar limb. 

Assuming mostly vertical magnetic field orientation to the solar surface,
the observed velocity oscillatory power 
is to be composed of not only the longitudinal velocity 
oscillations $\langle v_{\parallel} \rangle$ 
(angle brackets stands for the average 
root mean square value over time of the quantity),
 and a transverse 
(Alfv\'enic-like) $\langle v_{\perp} \rangle $ component.
The perpendicular component $\langle v_{\perp} \rangle$ is of 
special interest for constraining coronal 
heating models, as those waves are
expected to propagate readily
throughout the chromosphere and the into the corona
\citep{2005ApJS..156..265C}.
In the case of perpendicular to the solar surface observations,
close to disc center as in the case of FOV 2, 
we will detect only 
$\langle v_{\parallel} \rangle$ as the Doppler velocity.
Knowing $\langle v_{\parallel} \rangle$ from the observations of 
FOV 2 (close to disc center), we can calculate $\langle v_{\perp} \rangle$ 
from the observed velocity 
oscillations $\langle v_{obs}^2(\theta) \rangle$ 
(at an incidence angle $\theta$)
in FOV 1. The observed velocity
oscillations $\langle v_{obs}^2(\theta) \rangle$ can
be written into their components as (assuming that 
$\langle v_{\parallel} \rangle$ and $\langle v_{\perp} \rangle$ 
are not correlated):

\begin{equation}
\begin{aligned}
\langle v_{obs}^2(\theta) \rangle = \langle v_{\parallel}^2 \rangle \cos^2{\theta}
+ \langle v_{\perp}^2 \rangle \sin^2{\theta} = \\
=  \langle v_{\parallel}^2 \rangle \mu^2 +
\langle v_{\perp}^2 \rangle (1-\mu^2) 
\label{eqn:v_rms_from_angle}
\end{aligned}
\end{equation}
where $\mu = \cos{\theta}$, the cosine of the incidence angle of the
observation.

%

The average power spectrum of the \ion{Ca}{2} 854.2~nm
FOV 1 velocity (shown in Figure~\ref{fig:FOV_1_power})
exhibits lower oscillatory 
power compared to FOV 2 and has higher white-noise floor
compared to FOV 2, 
where the averaged FOV 1 data are shown as the black dots
in Figure~\ref{fig:average_PSD}. 

This is further 
illustrated in Figure~\ref{fig:FOV_power_histograms} where
the velocity fluctuation power in FOV 1 and 2 are compared.  
The red curves in Figure~\ref{fig:FOV_power_histograms} show the 
FOV 1 velocity fluctuation power between 5 and 20 mHz
for the different solar regions as segmented in
Figure~\ref{fig:FOV_1_mask}; the yellow curves show the
distributions for the solar regions for FOV 2.
Under the assumption that the velocities detected in FOV 2 are all due to longitudinal displacements 
(since the observed region was close to disc center at $\mu=0.98$),
%
we projected the observed distributions to what would be observed at an inclination angle of
$\mu=0.41$ by proportionally reducing the velocity magnitudes (green curves).
%
For all three types of solar structures (fibrils, network and plage),
the amount of oscillatory power in the projected FOV 2 
distributions is smaller than what is actually observed at that inclination in the FOV 1 data. 
%
Therefore, we believe that in FOV 1 we are observing not just projected vertical velocities ($v_{\parallel}$), but also an additional transverse ($v_{\perp}$) component. 
%

\begin{table}[]
  \begin{center}
\begin{tabular}{| c | c | c|}

Solar region & $\langle v_{\parallel} \rangle$ [km/s] 
& $\langle v_{\perp} \rangle$ [km/s]\\ \hline
Plage & 0.51 & 0.11 \\
Network & 0.48 & 0.17 \\
Fibrils & 0.45 & 0.18 \\

\end{tabular}

\caption{Velocity oscillation components (parallel 
         and perpendicular) inferred from the \ion{Ca}{2} 854.2~nm data from both 
         FOV 1 and 2.}
\label{table:oscillation_components}
  \end{center}
  \end{table}

Based on Equation~\ref{eqn:v_rms_from_angle} and taking
the median of the 
velocity oscillation distributions in 
Figure~\ref{fig:FOV_power_histograms} as
representative of the averaged oscillation power, we calculate the
magnitude of the transverse oscillations
$\langle v_{\perp}^2 \rangle$ for the different solar regions.
The resulting velocity components 
for the different solar regions are presented in 
Table~\ref{table:oscillation_components}. We can see that the plage region
has the highest longitudinal oscillations and lowest transverse components. The 
network and fibrils have similar values for the transverse oscillation
power, while the network has slightly higher longitudinal oscillation power.
On average, the value of transverse velocity component
$\langle v_{\perp}\rangle$
is a few times smaller than the longitudinal one 
$\langle v_{\parallel} \rangle$. This has been suggested in
previous modeling work (for 
example~\cite{2007ApJS..171..520C}).

Further observational studies could make use of samples at more
values of $\mu$ to better confirm and constrain the contributions 
of the two components. In addition, a more detailed treatment of 
radiative transfer effects at inclined viewing angles should be 
made.
Observations of more homogeneous solar regions (such as 
quiet Sun or network) would provide more assurance that 
similar structures were being compared and provide more accurate 
estimates for $\langle v_{\perp}^2 \rangle$. 

 \section{Modeling the observational signatures of acoustic waves in the chromosphere}
 \label{sec:Transmission_function}

 Optically thick diagnostics, such as the
 H$\alpha$ and \ion{Ca}{2} 854.2 nm 
 lines or the millimeter continuum
 used in this study,
 sample a wide range of heights in the chromosphere.
 Further, the interval sampled by these diagnostics are 
 changing dynamically during the continuous evolution of the chromospheric
 properties which are significantly out of thermodynamical
 equilibrium \citep{2019ARA&A..57..189C}.

The observed Doppler velocity response of a longitudinal 
wave-like perturbation in 
the solar atmosphere, with a wavelength
comparable to the extent of formation region of the spectral line, 
will be attenuated due to the 
inherent mixing of signals from different phases of the oscillatory fluctuation
\citep{1980A&A...84...96M}.
Modeling is needed to estimate the extent of this wave signature attenuation
and infer the true wave flux corresponding to the measured oscillatory amplitude.
%
%


 \subsection{Propagation of acoustic waves in the solar atmosphere}
 \label{subsec:Propagation_high_freq_waves}

 The dispersion relation for acoustic waves in the solar atmosphere is (following the
 derivation in \citealp{1974soch.book.....B}):
 \begin{equation}
 k_z^2 = \left ( \omega^2 - \omega^2_{ac} \right ) \frac{1}{c_s^2} -  \left ( \omega^2 - \omega^2_{BV} \right ) \frac{k_h}{\omega^2}
 \label{eqn:dispersion_relation}
 \end{equation}
 where $\omega = 2\pi \nu$ is the angular frequency;
 $k_h$ and $k_z$ are the horizontal and vertical wave numbers;
 $\omega_{ac} = \gamma g / ( 2 c_s^2)$ is the cutoff frequency
 below which no acoustic waves can propagate in the atmosphere.
 Under typical chromospheric conditions, the speed of sound is
 $c_s = 7$ km/s and $\nu_{ac} = 5.2$ mHz for
 an adiabatic index of  $\gamma = 5/3$;
 $\omega_{BV} = \sqrt{\gamma - 1} ~g / c_s^2$ is the Brunt-V\"{a}is\"{a}l\"{a}
  frequency which determines the lower cutoff frequency below which 
  gravity waves can propagate.

 The energy flux $F_{ac}$ carried by an acoustic wave 
 with a velocity amplitude squared per unit frequency
 $P(v^2_w)$ in the frequency interval of $\nu_0$ to $\nu_1$ is:

 \begin{equation}
 F_{ac} = \rho(z) \int_{\nu_0}^{\nu_1} v_{gr}(\nu')
 P(v^2_w(\nu')) \mathrm{d}\nu'
 \label{eqn:Acoustic_power}
 \end{equation}
 where $\rho(z)$ is the density at height \textit{z},
 $v_{gr}(\nu)$ is the group velocity. The group (propagation)
 velocity is the derivative $\partial \omega / \partial k_z$ from
 Equation~(\ref{eqn:dispersion_relation}) and in our case 
 it equals the sound speed times a factor of order unity above the
 acoustic cutoff frequency and zero below it: 
 
 \begin{equation}
     v_{gr}(\omega) = c_s \frac{\sqrt{\omega^2 - \omega_{ac}^2}}{\omega}
     \label{eqn:v_group}
 \end{equation}
 Compressive waves will also show a temperature
 fluctuation, which is proportional to the energy flux of the wave. 
 We can substitute in Equation~(\ref{eqn:Acoustic_power}) the wave 
 velocity fluctuation $v_w$ with the magnitude of the wave
 temperature fluctuation \citep{1974soch.book.....B}:
\begin{equation}
    v_w = \frac{c_s}{\Gamma_1 - 1}\frac{\delta T}{T_0}
  \label{eqn:temp_fluctuations}
\end{equation}
where $\Gamma_1$ is the adiabatic exponent describing how pressure 
responds to compression; $\delta T$ is the wave temperature 
fluctuation and $T_0$ is the mean temperature of the atmosphere. 

Based on 
Equation~(\ref{eqn:temp_fluctuations}), we can substitute in 
Equation~(\ref{eqn:Acoustic_power}) to get the final expression for the 
wave energy flux expressed via the relative
temperature fluctuations squared per 
unit frequency $P(\delta T / T_0)$: 

\begin{equation}
    F_{ac} = \frac{\rho(z) c_s^2}{(\Gamma_1 - 1)^2}
    \int_{\nu_{ac}}^{\nu_1} v_{gr}(\nu')
    P \left ( \frac{\delta T}{T_0} \right )
    \mathrm{d}{\nu'}
    \label{eqn:Acoustic_flux_temperature}
\end{equation}
where we integrate again over the frequency interval above the
acoustic cutoff and below the frequency at which the white noise floor 
dominates our signal.

ALMA is an ideal tool for the detection of the wave temperature fluctuations 
as it measures directly and linearly the plasma 
temperature in the chromosphere, with the observed 
brightness temperature equal to the
contribution-function-weighted mean temperature over 
the formation interval. This is a supplementary measurement
to the direct velocity measurement with IBIS in the 
\ion{Ca}{2} 854.2 nm line.

\subsection{Synthesizing wave observables with RADYN}
\label{subsec:Transmission_coeff_RADYN}


To model the line formation in the presence of waves 
we need time-dependent 
models of the solar atmosphere which incorporate all of the necessary 
physical processes of wave propagation (including
the optically thick radiative transfer effects).
We used the RADYN radiative hydrodynamic
code (RHD) \citep{1992ApJ...397L..59C, 1995ApJ...440L..29C, 1997ApJ...481..500C, 2002ApJ...572..626C, 1999ApJ...521..906A, 2015ApJ...809..104A}
to self-consistently model the propagation of high
frequency acoustic waves in the chromosphere. 
This 1D code solves the hydrodynamic equations coupled with
the radiative transfer equations
in non-local thermodynamical equilibrium
(NLTE). RADYN supports subphotospheric velocity drivers
defined by the user and treat time-dependent
NLTE ionization of the primary atomic species (H, He and Ca).
RADYN performs the radiative transfer calculations to generate time-dependent synthetic line profiles consistent with the wave dynamics.
Based on these synthetic observables, we are
able to interpret our data in terms of the realistic
solar plasma parameters
and estimate the amount of flux carried 
by the acoustic waves in the chromosphere.

 In our numerical setup, we use a 191 point atmosphere with
 6-level atom models for Hydrogen and Calcium and
 9-level atom for He -- including singly and doubly ionized states. 
 We use an open (transmitting) upper boundary condition
 where the corona is maintained at constant temperature
 1 MK at height of 12 Mm and a bottom piston boundary condition at fixed
 temperature of 5944 K. The piston bottom boundary allows for driving
 vertically propagating acoustic waves with arbitrary properties 
 which are defined by the user.

\begin{figure}
	\includegraphics[width=.45\textwidth]{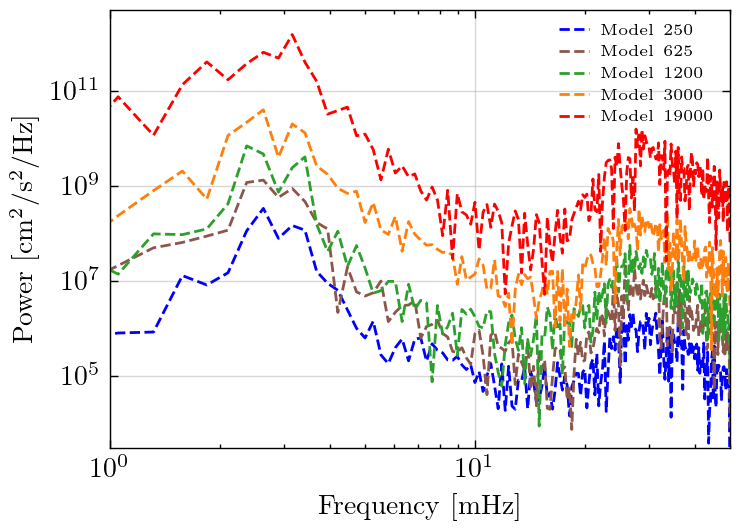}
	\caption{The vertical velocity power spectrum of the bottom boundary condition of the 
		different models we ran. The overall shape of the PSD is taken from 
		\cite{2006ApJ...646..579F} and scaled by a multiplicative factor and then 
	supplied as a bottom boundary (vertical velocity) condition in the RADYN runs.
    The PSDs presented here are the Fourier transforms of the actual 3 second 
     sampled RADYN run bottom boundary velocity.}
	\label{fig:RADYN_drivers}
\end{figure}

\begin{figure}
	\includegraphics[width=.45\textwidth]{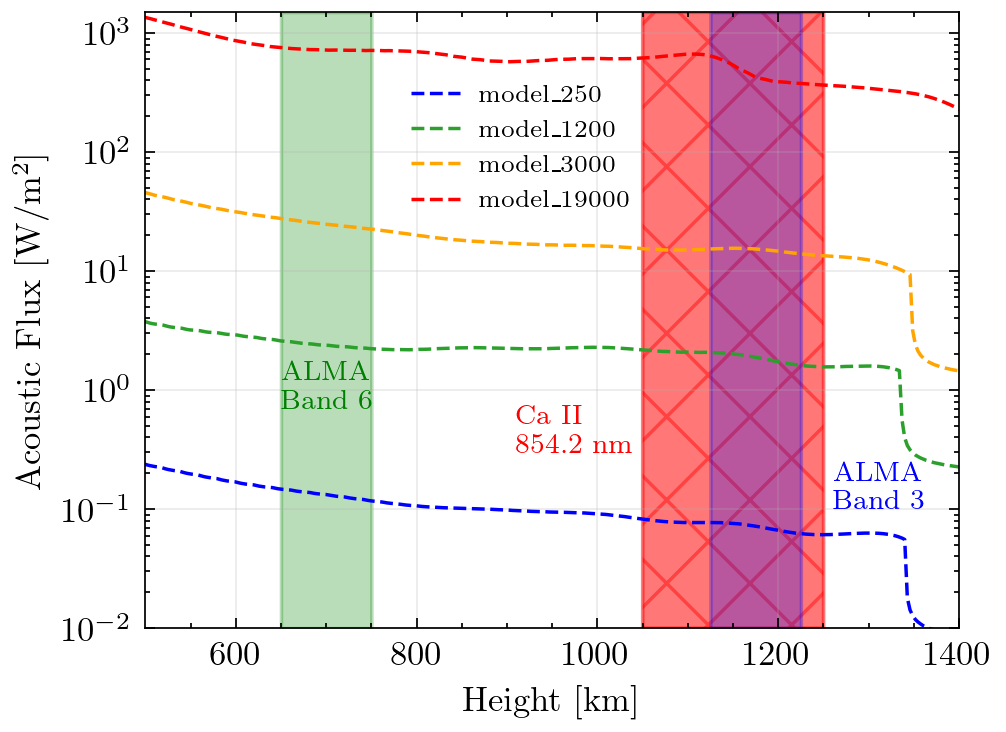}
	\caption{Acoustic flux dependence on height in
	the RADYN models, described in
		Section~\ref{subsec:Transmission_coeff_RADYN}. The shaded regions show the formation
		regions in the atmosphere for different spectral diagnostics.}
	\label{fig:RADYN_ac_chromosphere}
\end{figure}

 We ran a grid of models with an increasing amount of wave energy being
 injected through the bottom boundary condition. These models are used 
 consistently throughout
 the paper with the name format of \emph{model\_XXXX}, where the increasing
 numerical factor stands for a stronger bottom boundary driver.
 We used the bottom boundary vertical velocity drivers 
 presented in \cite{2006ApJ...646..579F}, 
 scaling their amplitude by a multiplicative factor
 to achieve the desired amount of input wave energy. 
These drivers specify a power spectrum of 
sub-photospheric velocities, at a range of oscillatory frequencies 
from 1 to 50 mHz, that have different relative amplitudes.
 To synthesize the velocity time series
 for the bottom boundary condition, we used the
 power spectrum of the driver initialized with random phases.
 The vertical velocity PSDs of the used bottom boundary drivers are
 presented in Figure~\ref{fig:RADYN_drivers}. 
 The resulting acoustic flux in the chromosphere
 from those runs are presented in 
 Figure~\ref{fig:RADYN_ac_chromosphere}. We have
 calculated the acoustic fluxes 
 from Equation~(\ref{eqn:Acoustic_power}), where we have 
 interpolated the models on a 4~000 point
 equidistant height grid to remove the 
 movement of the grid points during shock passages. To compute the average 
 acoustic fluxes, for each grid point we 
 calculate the mean of the sound speed and the plasma density 
 and filter the vertical velocity 
 between 5 and 50 mHz.  
 We have chosen time series starting about 
 10 minutes after the initialization of the simulation, 
 when a steady state is reached. We have further excluded models
 with higher acoustic fluxes above \emph{model\_3000} as the
 \ion{Ca}{2} 854.2 nm line core goes into emission and do not 
 reproduce the observations.

To compare the high-frequency signal in the synthetic 
observables and
in the IBIS observations we 
took the simulated \ion{Ca}{2} 854.2 nm line profiles from the
RADYN runs at 0.5 second intervals
and averaged
them to the temporal and spectral sampling of the IBIS instrument. 
We did apply the spectral PSF of IBIS, but this did not 
change the final results significantly as its spectral
resolution is very high 
($R >$~200,000) for the \ion{Ca}{2} 854.2~nm line. To measure the 
synthetic velocities, we used the same 
methods (i.e. line center fitting) used for the real
data processing.

\subsection{Estimating the acoustic flux in the \ion{Ca}{2} 854.2~nm
	        data based on RADYN}
\label{subsec:RADYN_T_CaII}	        

Given the attenuation of the wave amplitude discussed in
Section~\ref{subsec:Propagation_high_freq_waves}
we need to find a relationship
between the observed Doppler velocities, $v_{obs}$,
and the true wave amplitudes $v_w$.
One approach would be to discretize the acoustic flux in the observed bins
and then correct it for the wave amplitude attenuation
for each frequency bin separately (as in \cite{2009A&A...508..941B}):
 
 \begin{equation}
 F_{ac} = \rho(z) \sum_{\nu_i=\nu_{ac}}^{\nu_1} v_{gr}(\nu_i) 
 \langle v^2_{obs}(\nu_i) \rangle / {\cal{T}}^2(\nu_i)
 \label{eqn:Acoustic_power_approx}
 \end{equation}
 where the coefficient ${\cal T}(\nu_i)$ quantifies the attenuation of
 the wave velocity amplitude at frequency $\nu_i$ 
 due to the extended formation height of the
 spectral line. In this case the quantity
 $\langle v^2_{obs}(\nu_i) \rangle$ is the
 observed amount of velocity oscillation
 power per frequency bin and has units of velocity squared.
 ${\cal T}(\nu_i)$ is defined as the ratio of the 
 observed wave amplitude to the real (physical) wave amplitude
 in the middle of the formation region of the spectral line. 
 We measure $\langle v^2_{obs}(\nu_i) \rangle$ from the Doppler velocities in our observations, and estimate $\rho(z)$, $v_{gr}$ and 
 $\cal T$ from the self-consistent RADYN simulations.


We derive $\cal{T}$ by computing the square root of the ratio between 
the Doppler velocity power in the synthesized \ion{Ca}{2} 854.2~nm line
(Panel c) in Figure~\ref{fig:ALMA_RADYN}
and the power of the actual velocity
field in the RADYN atmosphere at the formation height of the spectral diagnostic for each frequency bin (Panel b) of Figure~\ref{fig:ALMA_RADYN}.
For the effective formation height of \ion{Ca}{2} 854.2 nm,
we chose the peak of the velocity response 
function.
For our models, presented in Figure~\ref{fig:ALMA_RADYN}, 
the peak of the \ion{Ca}{2} 854.2 Doppler velocity 
response function was at 1150 km, which is at about the same height
as the average $\tau=1$ surface for the line core. The extent of the 
velocity response function is shaded as the red region in 
Figure~\ref{fig:RADYN_ac_chromosphere}.
We use 5 mHz frequency averaging windows 
as the coefficient $\cal T$ and the power does not 
change significantly over this region and the averaging removes
the inherent uncertainty in the power spectra.
The inferred transmission coefficient $\cal T$ for 
\ion{Ca}{2} 854.2 nm is presented in Panel d) of
Figure~\ref{fig:ALMA_RADYN}. The Doppler signal of the
low frequency waves (below 10 mHz) is
less attenuated as their wavelength is significantly larger than the 
formation layer of the spectral diagnostic.
However, the high frequency wave signal is  
significantly attenuated, as shown before
\citep{2009A&A...508..941B}.

We note that our RADYN simulations can generate a high-frequency 
velocity signal due to the steepening of the acoustic waves 
propagating from the photosphere. The waves steepening into
shocks in the chromosphere 
have saw-tooth-like velocity profile, which when Fourier decomposed creates 
power law tails that extend to significantly higher frequencies than that 
of the driving wave \citep{2009A&A...494..269V}.
The modeling approaches in the previous studies 
\citep[for example][]{2009A&A...508..941B} using 
monochromatic high-frequency waves and static 1D atmospheric models
overestimate the transmission coefficient 
$\cal T$ (and the inferred wave flux respectively)
at higher frequencies, 
as suggested by our estimation of $\cal T$ from the RADYN 
simulations. This is due to the fact in our dynamic 
solar atmosphere models the propagating wave 
packets are not monochromatic and their steepening creates
high frequency signal. 

\subsection{Estimating the acoustic flux in the ALMA data
	        based on RADYN}
\label{subsec:ALMA_RADYN_observables}

There are two differences between the temperature fluctuations 
observed with ALMA and the ones in the right hand side of  
Equation~(\ref{eqn:Acoustic_flux_temperature}). First and foremost,
the temperature response of ALMA will be a result of the wave fluctuations
convolved with the atmospheric response function.
Another complication, that makes Equation~(\ref{eqn:Acoustic_flux_temperature})
applicable with limited validity in its current form is the  
fluctuating height of formation of the ALMA 
continuum~\citep{2019ApJ...881...99M}, which makes 
using a particular
height of formation (and local plasma density at that height)
nonphysical.

To take into account of all the effects described in the 
the previous paragraph, we rewrite
Equation~(\ref{eqn:Acoustic_flux_temperature})
as:

\begin{equation}
    F_{ac} = \sum_{\nu'=\nu_{ac}}^{\nu_{max}}
    \mathcal{C}_{ALMA}(\nu')
    \left \langle \left ( \frac{\delta T}{T_0} \right )^2 \right \rangle
    \label{eqn:ALMA_RADYN_flux}
\end{equation}
where the proportionality coefficient, called from now on 
the ALMA transmission coefficient, 
$\mathcal{C}_{ALMA}(\nu)$ encapsulates 
the local plasma properties at the formation height of the ALMA 
radiation. We note that the $\mathcal{C}_{ALMA}(\nu)$ coefficient is
the proportionality coefficient between the brightness temperature 
fluctuation power and the wave energy flux in the atmosphere, 
whereas the attenuation coefficient $\mathcal{T}$ in
Section~\ref{subsec:RADYN_T_CaII} is the attenuation of the 
observed wave velocity amplitude and goes in the denominator 
of Equation~(\ref{eqn:Acoustic_power_approx}).

By using the RADYN atmosphere models to synthesize synthetic ALMA 
observables and compare them with the observations we take
into account those two effects as described below. We use the 
RH code~\citep{2001ApJ...557..389U} to synthesize the millimeter 
continuum from the RADYN model output. We used the RADYN atmospheric models 
(including the instantaneous electron densities and hydrogen level
populations) as an input for the RH code to
synthesize the mm-wave radiation 
corresponding to Band 3 (100~GHz/3 mm) 
and Band 6 (240~GHz/1.25 mm). The RH code takes into account the 
opacity from neutral hydrogen as well as H$^-$ and H$_2^-$, 
which are the main sources of opacity in the solar atmosphere
at millimeter wavelengths~\citep{Zlotnik_1968}. We averaged
the output intensity from the 0.5 second time steps of the RADYN
runs to the 2 second cadence of our ALMA observations. 

\begin{figure*}
	\begin{center}
 	\includegraphics[width=0.95\textwidth]{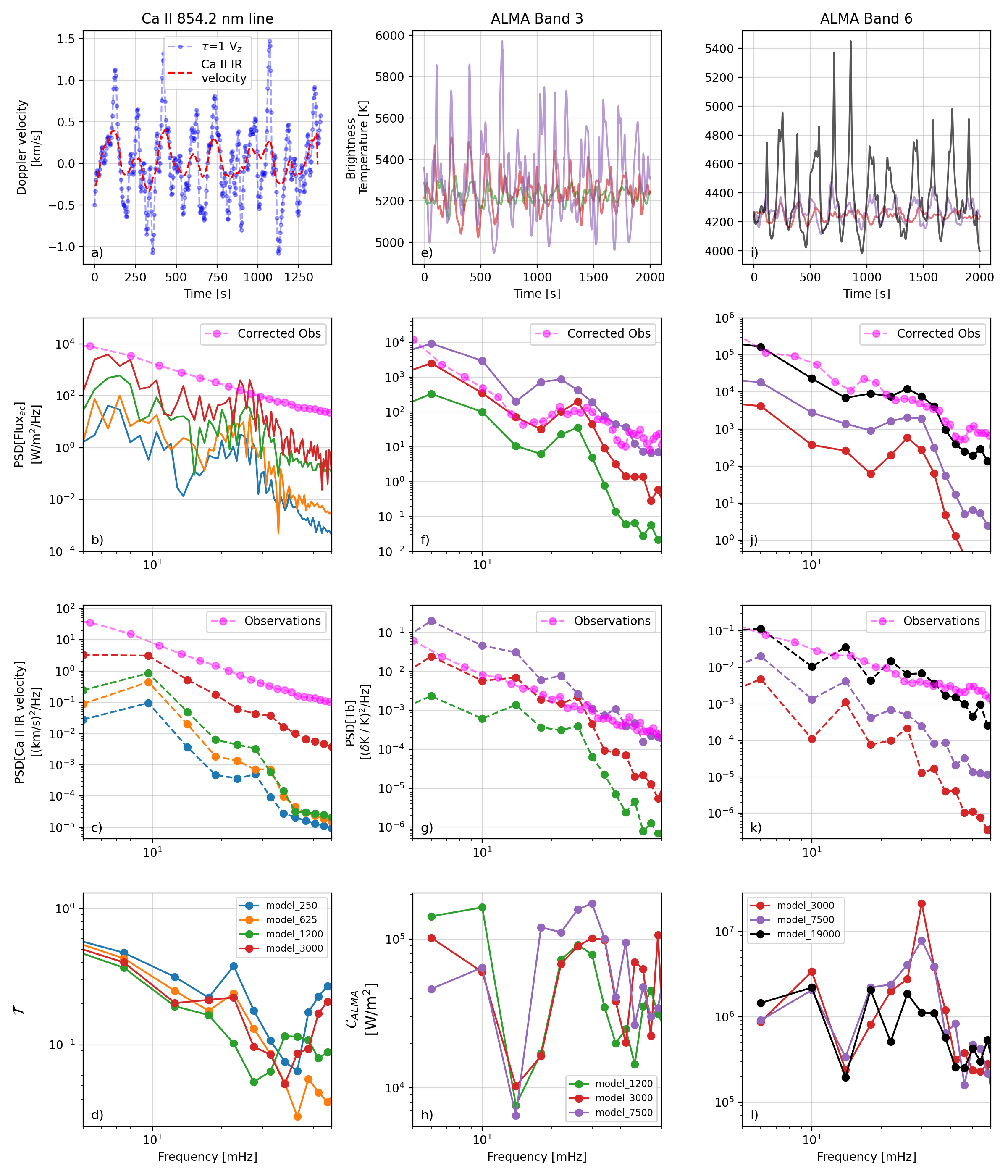}
 	\caption{Synthesized \ion{Ca}{2} 854.2 nm and 
 	ALMA responses to acoustic waves propagating
 	in the RADYN models for varying wave fluxes.
 		\textbf{\underline{Left column:}} results for \ion{Ca}{2} 854.2 nm. 
 		\emph{Panel a):} Vertical velocity (blue) in the RADYN 
 		simulation \emph{model\_3000} at the formation height of 
 		the \ion{Ca}{2} 854.2 nm line and the synthetic Doppler line 
 	    core
 		velocity (red) for the 
 		synthetic \ion{Ca}{2} 854.2 nm line from the same RADYN run. 
 		\emph{Panel b):} The PSD of the averaged acoustic flux at the formation 
 		height of the \ion{Ca}{2} 854.2 nm line. The magenta colored data are the 
 		averaged observed \ion{Ca}{2} 854.2 nm Doppler velocity PSD corrected with
 		the $\mathcal{T}$ in panel d).
 		\emph{Panel c):} PSD of the synthetic \ion{Ca}{2} 854.2 nm Doppler velocity. The 
 		magenta points are real observations.
 		\emph{Panel d):} $\mathcal{T}$ coefficient for 
 		the different RADYN models.
 		\textbf{\underline{Middle column:}} results for Band 3 (3 mm); 
 		\textbf{\underline{Right column}} results for Band 6 (1.25 mm). 
 		The panels in these columns
 		share the same information for the two ALMA Bands:
 		\emph{Panels e) and i):} synthesized ALMA brightness temperature from the RADYN models;
 		\emph{Panels f) and j):} Similar to Panel b), 
 		 but for the formation height of ALMA Bands 3 and 6;
 		 \emph{Panels g) and k):} Similar to Panel c, but for 
 		 the observed brightness temperature fluctuations in ALMA Bands 
 		 3 and 6;
 		 \emph{Panels h) and l):} The $\mathcal{C}_{ALMA}$ coefficient for
 		the different RADYN models.
 	    The color coding throughout this figure is consistent and corresponds to the 
 	    same models presented in the legend of the bottom row panels.}
 	\label{fig:ALMA_RADYN}
 	\end{center}
\end{figure*}

The extent of heights at which the ALMA 
Bands sample the atmospheric plasma temperature are shaded in 
Figure~\ref{fig:RADYN_ac_chromosphere} as the green region for Band 6
and the blue region for Band 3. We have determined those regions
as the height of $\tau = 1$ surface for the millimeter continuum
from the RADYN run most closely reproducing the observations. For 
Band 6 this was \emph{model\_19000} and for Band 3 is 
\emph{model\_3000} as shown in Figure~\ref{fig:ALMA_RADYN}.
The mean
formation height for the ALMA Band 6 is at 700 km and for Band 3
is 1150 km, while the physical widths of the formation regions 
shown in Figure~\ref{fig:RADYN_ac_chromosphere} correspond the
variation of the $\tau=1$ height in our models. Those heights 
are lower than the ones previously presented in the literature 
based on modeling \citep{2019ApJ...881...99M, 2020ApJ...891L...8M}
and observational \citep{2020A&A...634A..86P} methods.

The middle and right columns of 
Figure~\ref{fig:ALMA_RADYN} 
present the results from the ALMA 
spectral synthesis with the RH code
from the RADYN simulations and the resulting $\mathcal{C}_{ALMA}(\nu)$
coefficient. The left 
column corresponds to ALMA Band 3 (3 mm continuum) and the right
column corresponds to ALMA Band 6 (1.25 mm continuum).
Panels e) and i) of Figure~\ref{fig:ALMA_RADYN} show the temporal variation of the synthesized 
brightness temperature in the RADYN runs, which
agree well with previous studies of the 
millimeter continuum based on RADYN simulations \citep{2004A&A...419..747L, 2020A&A...644A.152E}. However, we do note that
the average synthetic temperatures of the RADYN
models (4250/5250 K for Band 3 and 6 respectively) 
are significantly lower than the observed ones
(7000/8500 K for Band 6 and 3)
in Figure~\ref{fig:ALMA_FOV}.
%
Panels f) and j) 
show the acoustic flux PSD 
at the average height of formation of the millimeter radiation (700/1150 km for Band 6/3 respectively), 
with a clear correlation between the amount of acoustic flux and
the amplitude of the brightness temperature fluctuations.
%
This correlation is further demonstrated in panels g) and k) of Figure~\ref{fig:ALMA_RADYN}, where we present the PSD of the 
modeled brightness temperature fluctuations.
This correlation is not surprising, as compressive waves have temperature
perturbations and ALMA measures the plasma temperature in the chromosphere.

$\mathcal{C}_{ALMA}$ is presented panels h) and l) of 
Figure~\ref{fig:ALMA_RADYN} where we have calculated it as
the ratio of the acoustic flux density at the formation height of the 
millimeter radiation (panels f) and j)) and the power density of the 
relative temperature fluctuations (panels g) and k)). We have
averaged the transmission coefficient $\mathcal{C}_{ALMA}$ over 
5 mHz frequency windows to smooth out the inherent noise in the 
power spectra. The $\mathcal{C}_{ALMA}$ coefficient
converges to the same values for
the RADYN runs with parameters 
closest to our observations (models 3000-19000). 
The converging values for $\mathcal{C}_{ALMA}$ at different 
acoustic fluxes confirm that
our modelling approach is not strongly dependent 
on the wave amplitudes (Equations~(\ref{eqn:ALMA_RADYN_flux})) and evades the complications of using 
Equation~(\ref{eqn:Acoustic_flux_temperature}) directly
for 
estimating the wave flux.

Due to the spatial smearing from the finite PSF of the ALMA beam
the observed oscillatory power is underestimated by 
a factor of 2 based on previous 
studies for Band 3
~\citep{2006A&A...456..713L,2015A&A...575A..15L,2020A&A...635A..71W}.
We do include this factor in our 
analysis only for Band 3, as we don't have an estimate for it for
Band 6.

\section{Inferred high frequency wave flux in the solar atmosphere}
\label{sec:RADYN_results}

\subsection{Wave flux inferred from the \ion{Ca}{2} 854.2~nm data}
\label{subsec:Ca_derived_flux}

\begin{figure}
	\includegraphics[width=.45\textwidth]{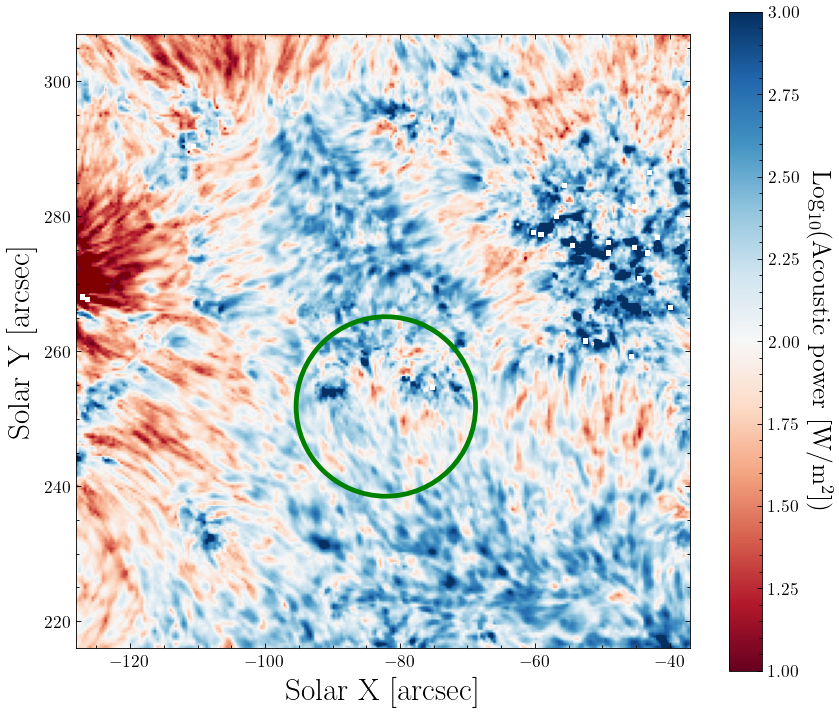}
	\caption{Acoustic flux inferred from the IBIS \ion{Ca}{2} 854.2 nm line 
		         observations for FOV 2 between 5 and 50 mHz. The calculation
		         uses the RADYN simulation results for the attenuation coefficient
		         in Section~\ref{subsec:Transmission_coeff_RADYN}. The green circle shows the 
		         FOV of ALMA Band 6 which was used
		         as an input for 
	     Figure~\ref{fig:Fluxes_comparisons}.}
	\label{fig:Ca_Acoustic_Flux}
\end{figure}

From the RADYN synthetic observables presented in 
the previous section, we had all the
constituents of Equation~(\ref{eqn:Acoustic_power_approx})
to compute the acoustic flux 
in our observations. The mean density at the mean formation height of the \ion{Ca}{2}
854.2 nm line is 5$\times$10$^{-9}$ kg~m$^{-3}$ 
(at 1150 km height in the model atmosphere).
This is comparable to estimates from previous
work~\citep{2020arXiv200802688A}. The values for
the attenuation coefficient $\cal T$ are presented in the bottom
left panel 
in Figure~\ref{fig:ALMA_RADYN} for the RADYN models 
with different driver strengths.
%
We used the values for $\cal T$ taken from the model 
producing the closest synthetic velocity/brightness temperature variations to the real data for the 
%
particular frequency bin and diagnostic.
This was the \emph{model\_3000} run for our IBIS observations.
This was also the
RADYN run with the highest piston amplitude that didn't cause the core of
the \ion{Ca}{2} 854.2 nm line to flatten out and go into emission (a condition not widely seen in our data).
%
We see that the attenuation 
 coefficients $\cal T$ from different model runs converge to similar values 
 at frequencies under 30 mHz. Above that, the attenuation coefficient $\cal T$
for different models do not agree that well, but this frequency region 
contributes relatively little to the total acoustic flux. 
Therefore, the exact driver strength does not appear to be important in the determination
of the attenuation coefficient and the derivation of the acoustic flux.

We computed the wave energy flux PSD in our observations 
based on our estimates for density and $\mathcal{T}$ from the RADYN simulations and 
the results are presented as the magenta data in the second row in
Figure~\ref{fig:ALMA_RADYN}. We can see that the energy flux
PSD derived from the observations is significantly higher 
in \ion{Ca}{2} observations compared to the simulations.
However, the shape of the wave flux frequency 
distribution is almost flat up to 20 mHz and resembles the one found in the RADYN models. 

At each pixel, we divide the observed
\ion{Ca}{2} 854.2 nm line core velocity power in each frequency bin by the corresponding value of $\cal T$ for 
\emph{model\_3000} and sum over the interval from 5 to 50 mHz to derive the total acoustic flux, with the results presented in Figure~\ref{fig:Ca_Acoustic_Flux} for the FOV 2 observations from 17:04 - 17:11 UT. 
%
We verified this result by calculating the same flux estimate using another FOV 2 dataset 
obtained from 15:39 to 15:46 UT, with very similar results.
%
This suggests that the derived acoustic fluxes are not significantly affected  by the seeing variations or 
evolution of the granular or supergranular conditions in the photosphere.
%

According to \cite{1977ARA&A..15..363W}, 
the average radiative losses in the middle (high)
chromosphere are ranging from about 2 kW~m$^{-2}$ 
(0.3 kW~m$^{-2}$) for the quiet internetwork to 
20 kW~m$^{-2}$ (2 kW~m$^{-2}$) for the plage regions.
The distribution of the acoustic flux in Figure~\ref{fig:Ca_Acoustic_Flux} 
suggests that over most of the FOV the acoustic flux can not be 
the dominant source of heating in the chromosphere. 
There are some disjoint regions with fluxes above 1 kW~m$^{-2}$ (shown in dark blue) where acoustic 
waves could be 
a significant source of chromospheric heating.
%
These regions of enhanced wave flux are located primarily
in the more fibril-free internetwork areas and also in the network and plage. 
The locations dominated by chromospheric fibrils (see Figure 2) almost uniformly have 
acoustic fluxes less than 0.1 kW~m$^{-2}$.
%

\begin{figure}
	\includegraphics[width=.45\textwidth]{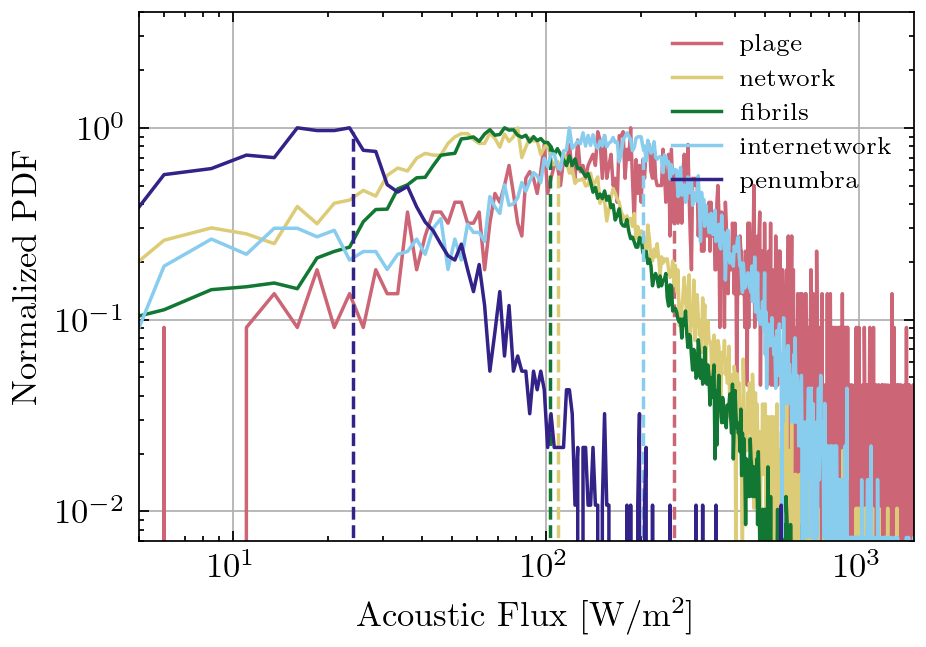}
	\caption{Histograms of the acoustic flux 
		in different regions of the solar surface for FOV 2. 
		The vertical lines show the median of the corresponding distribution. We 
		applied the mask in Figure~\ref{fig:masks_regions_of_sun} to the map of the
		acoustic flux in Figure~\ref{fig:Ca_Acoustic_Flux}.} 
	\label{fig:Flux_8542_ROS_distr}
\end{figure}

The distributions of observed acoustic flux
in the different solar regions (using the mask in Figure~\ref{fig:masks_regions_of_sun})
are presented in Figure~\ref{fig:Flux_8542_ROS_distr}. The averages and
the 10th/90th percentiles of the cumulative distribution are summarized 
in Table~\ref{table:ROS_flux_summary}.
The regions with the highest inferred acoustic flux are the
plage and internetwork regions. However, after further investigation 
of the shape of the spectral lines in the plage regions with highest
acoustic fluxes we found that the spectral line core fills in (flattens),
and the Doppler velocity derived from the line core minimum becomes less reliable. 
The line core of \ion{Ca}{2} 854.2 nm is determined by the local temperature
at the formation height of the line~\citep{2009A&A...503..577C}.
Strong impulsive heating in the plage regions can be responsible for this
dynamic peculiarity of the \ion{Ca}{2} 854.2 nm line profile. 
Therefore, we believe that the high velocity oscillation power observed 
at hotter (plage) locations might not be due to genuine wave motions but 
because of the the flattening of the line core.
Furthermore, the strong magnetic field in the plage regions might lead to MHD 
wave effects beyond the scope of this paper.
We observe a similar phenomenon in our RADYN
simulations, where if the wave driver is injecting too high velocity perturbations
(acoustic flux) we see the line shape of \ion{Ca}{2} 854.2 nm going into emission, 
when the shocks pass through the chromosphere. However, on average the 1D
simulations produce too narrow line core profiles  
\citep[as previously shown in][]{2009ApJ...694L.128L} which 
could be due to insufficient microturbulence 
in our simulation (RADYN has a default microturbulence of 2 km/s). 
Hence, we leave the detailed investigation of the dynamics of 
plage spectral line behavior for a future study.

Further analysis of the frequency dependence of the wave flux show that
about  60\% of the wave energy flux is in the 5--20 mHz frequency range, 30\% in the 20--40 mHz range and 10\% in the 40--60 mHz range. The pixels with significant relative 
contribution from the 20--40 mHz frequency range have lower
total acoustic flux and are mostly found in the plage and the network. 

In summary, our observations
show that acoustic wave dissipation is likely not the dominant heating 
mechanism for the middle chromosphere. However, acoustic waves could contribute
significantly to the quiescent state of the upper chromosphere
in the internetwork as the observed flux is of the order of magnitude 
of the radiative losses in that layer. Future work will constrain the wave flux
and dissipation in this layer using IRIS observations.

\subsection{Inferred wave flux from the ALMA observations}

Using the RADYN results presented in 
Section~\ref{subsec:ALMA_RADYN_observables} 
we are able to 
estimate the acoustic fluxes in the ALMA observations. We
match the observed brightness temperature fluctuations 
(the magenta points in panels g) and k) in Figure~\ref{fig:ALMA_RADYN})
to the closest RADYN model by the amount of brightness temperature 
RMS and then use the $\mathcal{C}_{ALMA}$ coefficient
from that RADYN run (bottom row in 
Figure~\ref{fig:ALMA_RADYN}) to calculate the amount of acoustic 
flux. To calculate the acoustic flux for Band 3 we used the observing block 
obtained between 17:31 and 17:41 UT and the
$\mathcal{C}_{ALMA}$ values for \emph{model\_3000}. 
For Band 6 we used the observing block between 16:03 and 16:11 UT 
and the $\mathcal{C}_{ALMA}$ values for \emph{model\_19000}.

\begin{figure*}
	\includegraphics{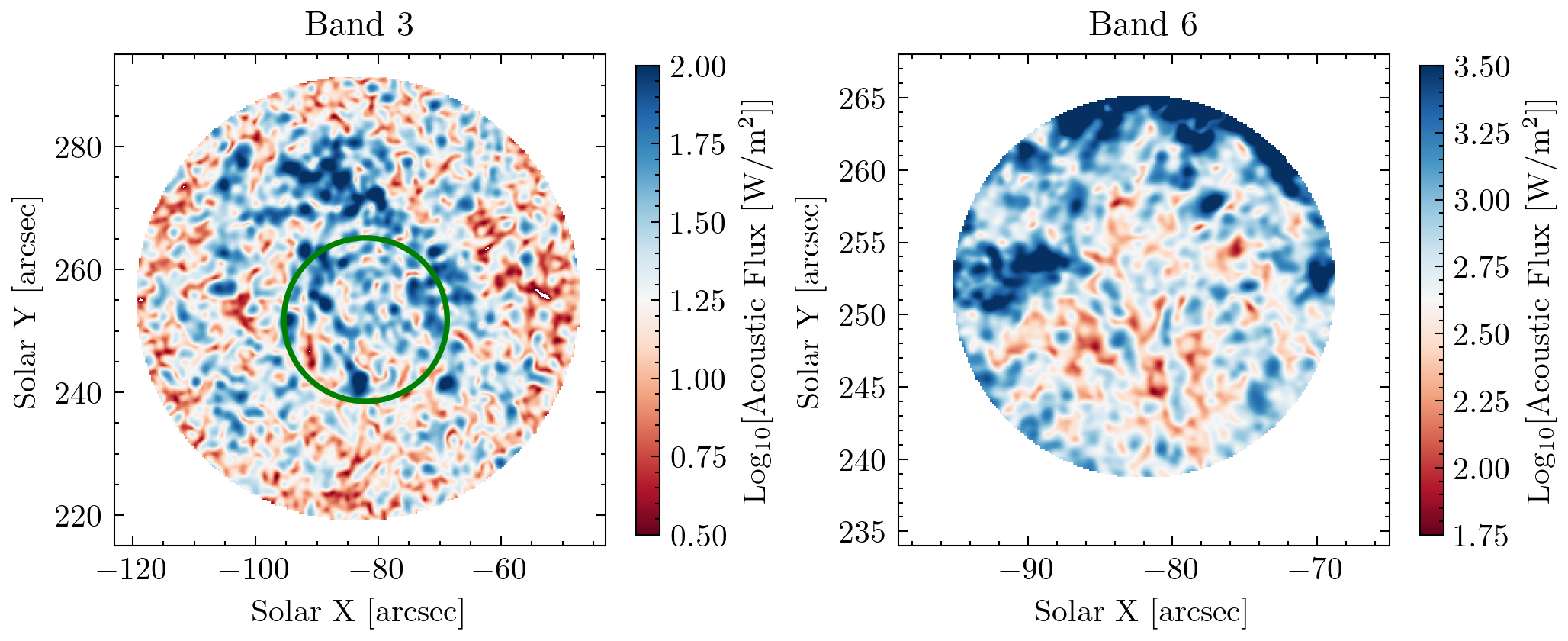}
	\caption{Acoustic flux derived from
		 ALMA Bands 3 (\emph{left panel}) and 
		 6 (\emph{right panel}) data based on the RADYN models. We 
	 have applied a circular mask to present only 
     the central part of the beam with the highest sensitivity and lowest
     synthesis noise. The green circle in the Band 3 FOV is the extent of 
     the Band 6 FOV.}
	\label{fig:ALMA_Acoustic_Flux}
\end{figure*}

The inferred acoustic flux for Bands 3
and 6 are presented in Figure~\ref{fig:ALMA_Acoustic_Flux}.
Circular masks were applied around the edges 
field of view as the noise 
outside of those regions are significant due to the 
decreasing sensitivity. We observe higher 
acoustic flux in Band 6 compared to Band 3, which is 
expected if the wave flux is
being dissipated as the waves propagate upward. The region of the FOV of
Band 6 is shown on the Band 3 FOV as the green circle. 
Since the two different ALMA Bands were observed
an hour apart,
they do agree to a certain extent but not fully as the solar 
atmosphere is changing on shorter time scales. Furthermore, the very 
limited FOV of Band 6 makes comparisons difficult.
Examining the frequency distribution of the wave flux
shows that the dominant source of signal in both ALMA Bands  
are the frequency range between 5 and 20 mHz, where 
more than 70\% of the signal is found. This agrees
with our results for the \ion{Ca}{2} 854.2 nm line, that the 
frequency range between 5 to 20 mHz contains most of the 
wave flux.

\begin{figure}
	\includegraphics[width=.45\textwidth]{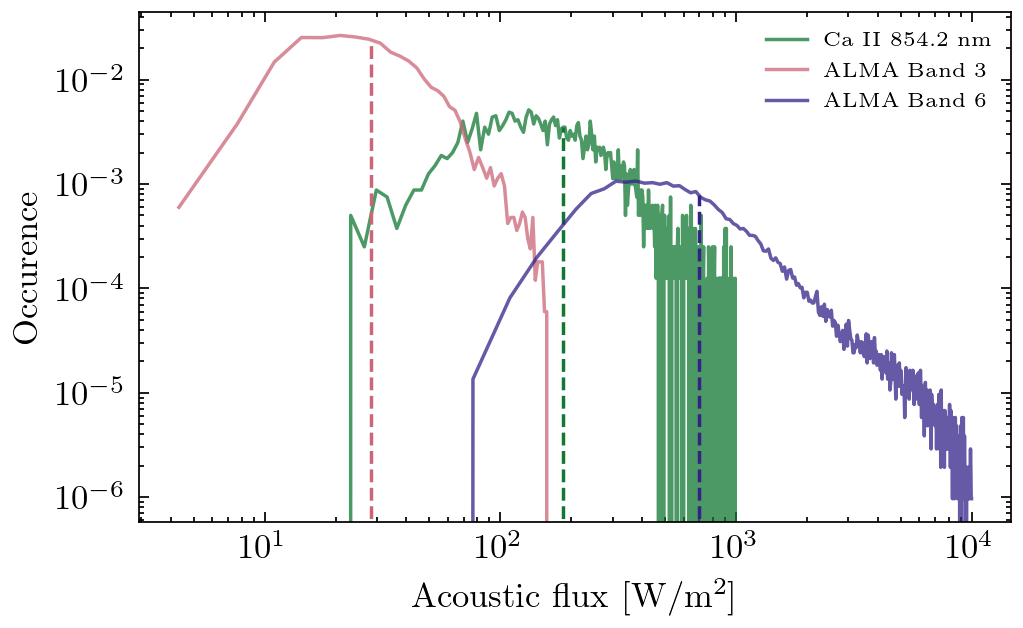}
	\caption{Acoustic flux distributions derived from the different diagnostics shown
		in the legend. The vertical lines are the medians of the distributions.}
	\label{fig:Fluxes_comparisons}
\end{figure}

To compare the results from IBIS and ALMA,
we compare the derived acoustic
flux over a common FOV corresponding to that of 
the Band 6 data. The distributions of the observed fluxes 
at those locations are presented in Figure~\ref{fig:Fluxes_comparisons}.
The ALMA Band 6 data exhibits the highest flux. They are followed by the 
\ion{Ca}{2} 854.2 nm and the ALMA
Band 3 data. This ordering of the amount of wave flux 
follows the height of formation of the diagnostics shown in Figure 
\ref{fig:RADYN_ac_chromosphere}.
Since Band 3 and 6 have similar systematics in 
their formation height we can estimate the 
dissipation between their formation heights. The average dissipated energy 
(flux difference) across the FOV of Band 6 data is around 
0.7 kW~m$^{-2}$. However, there is a significant high-power tail in the  
dissipated energy which is greater than 1 kW~m$^{-2}$ and is energetically
significant to maintain the quiet chromosphere in some confined regions. 
However, on average this is not enough 
to sustain the quiet middle chromosphere, 
being too small by an order of magnitude.  

\section{Conclusions and future work}
\label{sec:Conclusions} 

We obtained an extensive data set containing spectral 
observations covering 
from the upper photosphere and the middle 
(\ion{Ca}{2} 854.2 nm, H$\alpha$, and ALMA Band 6)
and upper 
chromosphere (ALMA Band 3). The rapid cadences of our 
observations
allow us to study high frequency dynamics of the chromosphere
in a rarely studied frequency regime.
Our observations have extended the work of \cite{2008ApJ...683L.207R}
to show that the power-law distribution of the Fourier PSDs 
is ubiquitous in all of the observed velocity and temperature diagnostics.
In particular, in this paper we have 
focused on the velocity diagnostics derived from \ion{Ca}{2} 854.2 nm, 
as tracers of upward propagating compressive waves, 
as well as brightness temperature fluctuations
from ALMA Band 3 (3 mm) and 6 (1.2 mm)
as indicators of local heating from those waves.
We found that the power law properties depend on the observed
solar region. We confirmed that the white-noise level in the power 
spectra is consistent with the photon shot noise in our data 
(see Appendix~\ref{App:Noise_floor}), indicating that seeing-induced crosstalk
or other sytematic noise is likely not responsible for
the detected Doppler velocity power.
The amplitude of the noise floor increases in more "active" regions of the field (e.g. plage), 
perhaps due to flatter line cores typical of these features.
%
Furthermore, the slope of the spectral diagnostics' 
power laws changes with 
the observed regions. ``Hotter'' regions (like network and plage)
exhibit less steep power law slopes as shown in 
Figure~\ref{fig:PSD_slope_histogram}. 
This flattening of the power law could be due to the filling of the core 
of the spectral profile of the \ion{Ca}{2} 854.2 nm line in the hotter regions 
(network and plage).
%
We believe the steeper power law 
is an important characteristic of the dynamic nature of the 
different solar surface features that has to be reproduced in future modeling efforts.

We compared the RMS values of velocity
oscillations observed at different
incidence angles ($\mu=0.98$ vs $\mu=0.41$). 
The RMS of the velocity 
fluctuations is significantly smaller at 
the higher incidence angle compared 
to the disc center observations, but still larger than
what would be expected if
the amplitudes of those fluctuations were only
due to the inclined viewing of vertically propagating
waves of the same magnitude as 
seen in our disk center observations.
Modeling the observed signal as a 
superposition of a longitudinal (vertical) component and a
transverse (possibly Alfv\'enic) component, we found that the transverse 
component had a RMS amplitude of about 0.15 km/s, compared to the 
vertical component amplitude of around 0.5 km/s. 

To characterize the acoustic waves that could explain the power laws 
in our observations we used
 the RADYN code to model propagation of waves from the upper convection zone into the chromosphere. 
 We used wave drivers (as bottom boundary conditions) 
 similar to the ones in~\cite{2006ApJ...646..579F} adjusted with a scaling factor. We
 ran the RHD models and then produced synthetic observables for both
 IBIS and ALMA, using the RADYN built-in radiative transfer 
 module in the former case and the RH code in the latter case.
 The dynamic RADYN models are able to
 reproduce the features of our observations in terms of total oscillatory power
 and slopes. Hence, we were able to correlate our observed oscillatory power in different diagnostics to the actual acoustic flux present at different heights in the simulations.

The acoustic flux derived from
the \ion{Ca}{2} 854.2 nm line
Doppler velocity data is estimated to be between 0.1 to 1
kW~m$^{-2}$. The lowest amount of flux is found in the penumbra 
and fibril regions and the highest in the internetwork and plage 
regions. We believe that the inclined nature of the 
magnetic field in the penumbra and the fibril region
plays role in the observed lower fluxes. 

Furthermore, as discussed in Section~\ref{subsec:Ca_derived_flux}, the high values in the plage region above 1 kW/m$^2$ require 
further examination
due to the changes in the Ca II 854.2 nm line profile in plage regions, 
which leads to spuriously high measured Doppler shifts.

Furthermore, we found that most of the contribution 
to the acoustic flux comes from the 5-20 mHz frequency interval (around 60\% from the total acoustic flux)
and that high frequency waves above 40 mHz do not contribute significantly (less than 10\% of the total
observed flux). 

We also compared the brightness temperature fluctuations 
in our ALMA data with the synthetic observables from the RADYN runs 
and inferred the acoustic flux by using the correlation between 
brightness temperature fluctuations and acoustic flux in the different RADYN model runs. 
Based on this comparison we can infer that
Band 6 has acoustic flux on average of 0.7~kW~m$^{-2}$, compared
to the formed higher in the atmosphere Band 3 
which contains about 0.03~kW~m$^{-2}$. From these two observations at different heights,
we can infer that the average wave flux dissipated between 
two layers probed by ALMA is about 0.7~kW~m$^{-2}$, which is not sufficient 
to heat the middle chromosphere, but is a significant 
contribution to its energy budget. This result agrees 
quantitatively with previous work by \cite{2020A&A...638A..62N},
who used ALMA to compute the heating in small scale 
chromospheric brightnenings.
However, in certain regions the
dissipated wave flux exceed the threshold of
2~kW~m$^{-2}$ which is sufficient 
to maintain the quiet chromosphere locally.
We believe that the limited 
spatial resolution of the ALMA observations could lead to 
an underestimation of the wave flux, compared to our \ion{Ca}{2}
data~\citep{2015A&A...575A..15L, 2020A&A...635A..71W} and further
observations with higher angular resolution (more sparse ALMA
configuration) will provide better constraints on the wave flux.

Another peculiarity between our observations
and ALMA data is the fact that the Band 3 corresponds most 
closely to \emph{model\_3000} while Band 6 to 
\emph{model\_19000}. \emph{Model\_19000} has twenty times
higher wave energy flux than \emph{model\_3000} in the chromosphere. This discrepancy could be due to different reasons -- either the heights of formation of the ALMA continuum 
are inaccurate in the RADYN models or there is a wave
dissipation mechanisms not included in the RADYN models.
If the height in the atmosphere where the ALMA
continuum originates from is determined inaccurately, this will lead to incorrect acoustic flux determination 
and result in differing RADYN models corresponding 
to the observations. The latter possibility of missing
physics is also very probable due the 1D hydrodynamic nature of 
the RADYN code which might be omitting 
the required physics to treat fully
all the relevant wave damping mechanisms.

Our work raises further questions such as what is the
role of the magnetic field on the wave propagation characteristics
in the solar atmosphere? Future observations with multiwavelength 
spectropolarimetric capabilities throughout the photosphere and 
the chromosphere
from Daniel K. Inouye Solar Telescope (DKIST)
\citep{2020SoPh..295..172R, 2021SoPh..296...70R}
will be able to address that. Also,
the higher throughput of the future generation solar telescopes will 
help with driving down the white noise floor and provide 
simultaneous spectral observations at different heights in the solar atmosphere. 
Another interesting aspect we will pursue in a following 
publication is the amount of velocity oscillations in the upper 
chromosphere and the transition region observed 
cotemporally with IRIS during our April 2017 campaign. 

The numerical side of our work requires further investigation as well. 
One important question that will be addressed in future work 
is the sensitivity of 
our results on the numerical setup that was
utilized. For example, how does 
the transmission coefficient $\mathcal{T}$ depend on the model atmosphere and the
number of grid points in the atmosphere?
Furthermore, studying
wave propagation in 3D is essential for understanding the observed signals as
the nature of the observed chromospheric structures is strongly
non-vertical and non-local~\citep{2019ARA&A..57..189C, 2021RSPTA.37900185E}.
Even though current 3D RMHD 
solar models have significantly differing wave propagation 
properties~\citep{2020arXiv200705847F}, 
further modeling with realistic solar atmospheres in three dimensions
is essential.

\acknowledgements
The authors would like to thank the observing staff at the Dunn Solar Telescope 
at Sacramento Peak for the support during our observing run in April 2017. The authors are indebted 
to Amanda Alexander and Gianna Cauzzi
for proofreading the manuscript and the anonymous referee whose 
suggestions improved the manuscript greatly.
Data in this publication were obtained with the facilities of the National Solar Observatory, which is operated 
by the Association of Universities for Research in Astronomy, Inc. (AURA), under cooperative agreement 
with the National Science Foundation.
IBIS has
been designed and constructed by the INAF/Osservatorio
Astrofisico di Arcetri with contributions from the Università di
Firenze, the Università di Roma Tor Vergata, and upgraded
with further contributions from NSO and Queens University
Belfast. MEM was supported for part of this work by the Hale Graduate 
Fellowship from the University of Colorado and by the DKIST Ambassador Program. Funding for the DKIST 
Ambassadors program is provided by the National Solar
Observatory, a facility of the National Science Foundation, operated under Cooperative Support Agreement
number AST-1400405.
Furthermore, this paper makes use of the following ALMA data: ADS/JAO.ALMA\#2016.1.01129.S. ALMA is 
a partnership of ESO (representing its member states), NSF (USA) and NINS (Japan), together with NRC 
(Canada), MOST and ASIAA (Taiwan), and KASI (Republic of Korea), in cooperation with the Republic of 
Chile. The Joint ALMA Observatory is operated by ESO, AUI/NRAO and NAOJ. The National Radio 
Astronomy Observatory is a facility of the National Science Foundation operated under cooperative 
agreement by Associated Universities, Inc.

\vspace{1mm}
\facilities{DST(IBIS), ALMA, IRIS}

\software {SolarSoft~\citep{2012ascl.soft08013F}; 
Matplotlib \citep{matplotlib}; 
NumPy \citep{numpy}; SciPy \citep{scipy}.
The scripts and the Jupyter notebooks
containing the code for this project are available on 
the public repository of the author:
\url{https://bitbucket.org/super_momo2/comps_code/src/master/}}. Paul Tol's excellent
color tables (\url{https://personal.sron.nl/~pault/}) were used for making the categorical
figures throughout the paper.

\bibliography{Sections/Bibliography.bib}{}

\appendix

\section{Estimating the noise floor in the power spectra of the observed Doppler velocities}
\label{App:Noise_floor}

The power spectra presented throughout this paper 
(for example in Figures \ref{fig:Power_spectra_overview}
and  \ref{fig:average_PSD})
exhibit white noise behavior at the high 
frequency limit. To interpret 
the derived properties from the observed 
power spectra correctly 
we need to understand the noise sources in our data that
could contribute to the wave signal. Spectral line
profiles with different shape would be affected to a different  
degree from photon noise due to their varying shapes and intensity 
levels. For example, deeper and narrower profiles
chromospheric profiles of the \ion{Ca}{2} IR lines would be less
susceptible to Doppler velocity measurement shot noise, compared
to plage or AR shallower spectral profiles 
whose cores fill up and flatten. Hence, we modeled the effect of
photon shot noise on 
each different solar surface feature described 
in Figure~\ref{fig:masks_regions_of_sun}.

To estimate the white noise floor properties due to photon shot 
noise to the \ion{Ca}{2} 854.2 nm line velocity power spectra 
we adopted a Monte-Carlo approach. We chose 150 random
spectral profiles from the each region in 
Figure~\ref{fig:masks_regions_of_sun}. We did not 
choose the average spectral line 
profile for each region of the Sun to 
be representative as averaging over space and time does not 
represent the instantaneous realization of the 
spectral profiles. We computed 1500 noise realizations 
for each chosen spectral line profile. 
The noise for each wavelength point was calculated by using the
ADU (2.5 e-/DN) of the camera (Andor iXon 885) used that day to calculate
the number of photons. Since the signal to noise ratio was 
on the order of a few hundred (even for the line cores),
we applied a Gaussian noise statistics 
to the estimated photon shot noise levels. The Doppler velocity from 
the simulated time series of 1500 noise realizations was measured 
with the same techniques used for reducing our IBIS data (described in 
Section~\ref{subsec:power_spectra_properties}). The measured
Doppler velocity power spectrum density
was white noise, as expected for uncorrelated noise.
The bottom panel of
Figure~\ref{fig:HF_noise_levels} presents the distribution of the
median noise level
in our estimation. The synthetic noise distributions match well with 
the observed ones. Hence, 
we can assume that seeing induced crosstalk
is not the dominant source for the high frequency 
white noise floor.

\end{document}